\documentclass[12pt]{article}
\usepackage{amsmath}
\usepackage{amssymb}
\begin{document}
\begin{center}
{\bf {\large{Constraints, Symmetry Transformations and Conserved  \\ Charges for Massless Abelian 3-Form Theory}}} 

\vskip 3.4cm

{\sf  B. Chauhan$^{(a)}$,  A. K. Rao$^{(a)}$, R. P. Malik$^{(a,b)}$}\\
$^{(a)}$ {\it Physics Department, Institute of Science,}\\
{\it Banaras Hindu University (BHU), Varanasi - 221 005, (U.P.), India}\\

\vskip 0.1cm

$^{(b)}$ {\it DST Centre for Interdisciplinary Mathematical Sciences,}\\
{\it Institute of Science, Banaras Hindu University, Varanasi - 221 005, India}\\
{\small {\sf {e-mails: bchauhan501gmail.com, amit.akrao@gmail.com;  rpmalik1995@gmail.com}}}
\end{center}

\vskip 2.0 cm

\noindent
{\bf Abstract:}
 We demonstrate  the existence of the first-class constraints on the massless Abelian 3-form  theory 
which generate the {\it classical} gauge symmetry transformations for this theory in any arbitrary D-dimension of spacetime.
We write down the explicit expression for the generator in terms of these first-class constraints.
Using the celebrated Noether theorem, corresponding to the gauge symmetry transformations,
we derive the Noether conserved current and conserved charge. The {\it latter} is connected with the first-class
constraints of the theory in a subtle manner as we demonstrate clearly in our present investigation.
We comment on the first-class constraints within the framework of Becchi-Rouet-Stora-Tyutin (BRST)
formalism where the conserved (anti-)BRST charges are the generalizations of the above {\it generator} for the
classical gauge symmetry transformation.  The standard Noether conserved (anti-)BRST charges are found to be non-nilpotent.
We derive the nilpotent versions of the (anti-)BRST charges.
One of the interesting observations of our present endeavor is the result that only the {\it nilpotent} versions
of the conserved (anti-)BRST charges lead to the annihilation  of the physical states by the operator form of the 
first-class constraints at the {\it quantum} level which is consistent with the Dirac quantization condition
for the systems that are endowed with any kind of constraints. 
We comment on the existence of the Curci-Ferrari (CF)  type restrictions from different theoretical angles, too.



\newpage
\section {Introduction}
One of the most successful theories in the realm of theoretical high energy physics (THEP) is the 
standard model of particle physics (SMPP) where the agreements  between theory and experiments are 
outlandish (see, e.g. [1-5] and references therein). This successful theory 
(describing three\footnote{The SMPP is primarily the quantum field-theoretic description of the strong, weak and 
electromagnetic interactions of nature. However, it does not say anything about the quantum theory of gravity. In
addition, this theory contains a large number of parameters which is 
{\it not} a decisive signature of a beautiful theory.} 
out of four fundamental interactions of nature)
is based on the quantum 
field theory of an interacting non-Abelian 1-form $(p = 1)$ gauge theory where the weakly interacting
neutrinos  are taken to be massless. However, the modern experiments have established conclusively that  
the neutrinos are {\it massive}. Hence, it is very clear that the theory of SMPP is {\it not} complete and one 
has to go beyond the realm of the validity of SMPP (as far as the theoretical aspects of the SMPP are concerned). 
At present, the superstring theories are the most promising candidates which are expected to provide
(i) the quantum theory of gravity, (ii) the complete unification of {\it all} the four fundamental interactions 
of nature, and (iii) the precise theoretical framework in THEP whose lower energy limit 
is the theoretical physics of SMPP (see, e.g. [6-10] and references therein). 
One of the key theoretical observations in the context of (super)string theories is the appearance 
of a tower of higher $p$-form $(p = 2, 3, ...)$ basic gauge fields in {\it their} quantum excitations.

Against the backdrop of the above paragraph, it is but natural to think about the possibility
of developing a quantum field theory that is based on the higher $p$-form ($ p = 2, 3,...$) gauge fields which provide
a theoretical set-up beyond the theoretical regime of the SMPP in THEP. Hence, the quantum field theory, associated with 
the  higher $p$-form ($p = 2, 3...$) gauge fields,   
is important and essential to be studied thoroughly because of its connections with the (super)string theories. 
The central objective  
of our present investigation is to study the constraint structures of  the {\it massless} 
Abelian 3-form gauge theory and
establish that they belong to the first-class variety in the 
terminology of Dirac's prescription for the classification scheme of constraints 
(see, e.g. [11-15] for details). The first-class constraints generate the {\it classical}
local, continuous and infinitesimal gauge symmetry transformation of  the above theory  as they are found to be 
present in the standard expression for the generator of this specific  gauge  
transformation (see, e.g. [16, 17]). We have derived the Noether conserved current and charge (corresponding to the {\it classical} 
local, continuous and infinitesimal gauge symmetry transformation) and established that the {\it latter} is  nothing but 
the standard definition of the generator in terms of the first-class constraints.

We have elevated the {\it classical} local, continuous and infinitesimal gauge symmetry transformation to the {\it quantum} level 
within the framework of Becchi-Rouet-Stora-Tyutin (BRST) formalism [18-21] and shown that traces of the first-class constraints of the 
{\it original} massless  Abelian 3-form theory  are present in the physicality criteria 
(i.e. $[Q^{(1)}_{(a)b}]| \, phys > = 0$) w.r.t.  the conserved and  nilpotent (anti-)BRST charges $[Q^{(1)}_{(a)b}]$  when we demand 
that the physical states $(|\, phys >)$, in the total quantum Hilbert space of states,
are {\it those} that are annihilated by the conserved and off-shell nilpotent (anti-)BRST charges
 $[Q^{(1)}_{(a)b}]$.  It has been explicitly demonstrated, in our present endeavor, that the
operator  form of the first-class constraints of the original theory annihilate the physical states (i.e. $|\, phys > $) of 
the {\it total} quantum Hilbert space of states within the framework of 
BRST formalism which is consistent with the Dirac quantization condition
(see, e.g. [11-15]) for theories endowed with any kind of constraints.  At this stage, it is pertinent to point out that, 
in the context of the massless 
D-dimensional Abelian 3-form theory, the standard Noether theorem leads to the derivations of the conserved (anti-)BRST charges 
[$Q_{(a)b}$] but they are found to be non-nilpotent. In our earlier work [22], we have developed a systematic method to derive the 
nilpotent versions of the conserved (anti-)BRST charges $[Q_{(a)b}^{(1)}]$ from the standard conserved 
but non-nilpotent (i.e. $Q_{(a)b}^2 \ne 0$)
Noether conserved (anti-)BRST charges.

In our present endeavor, we exploit this proposal [22]
to obtain the off-shell nilpotent versions of the conserved (anti-)BRST charges and demonstrate that the physically criteria 
w.r.t. these {\it latter} charges lead to the appearance of the first-class constraints (of the original starting theory) in their 
operator forms which annihilate the physical states $(|\, phys >)$ at the {\it quantum} level. This observation is consistent with the 
Dirac quantization condition for the theories that are endowed with any kind of constraints [11-15]. We lay emphasis on the fact that 
the standard Noether conserved (anti-)BRST charges ($Q_{(a)b}$) do {\it not} lead to the annihilation of the physical states
by the operator form of the first-class constraints of the {\it classical} D-dimensional massless Abelian 3-form theory. 
In addition, the non-nilpotent behavior of the (anti-)BRST charges creates problem in the precise discussion on the 
topic of BRST cohomology. We claim
here that this observation is {\it general} and it is true for any BRST-quantized theory which is endowed with the non-trivial CF-type
restriction(s). All the higher (non-)Abelian $p$-form ($ p = 2, 3,...$) gauge theories, 
discussed within the framework of BRST formalism, belong to this category.

Our present investigation is essential, interesting and important on the following counts. First of all, in all our previous
works related with the theory under consideration (see, e.g. [23-26] for details), we have {\it not} demonstrated the existence
of the {\it classical} local, continuous and infinitesimal gauge symmetry transformation in the terminology of the first-class
constraints in a  systematic manner. Second, our present BRST-quantized massless Abelian 3-form theory is endowed with {\it three}
CF-type redirections. We wish to demonstrate that the standard Noether conserved (anti-)BRST charges are non-nilpotent
in our present case, too, as we have shown in the context of the {\it modified} amssive and massless Abelian 2-form theories [22, 27].
We obtain the nilpotent versions of the (anti-)BRST charges for our present case following the proposal of our earlier work [22]. Third, our present 
endeavor is a modest
step towards our main goal of proving the 6D St$\ddot u$ckelberg-modified {\it massive} Abelian 3-form gauge theory to be a
massive field-theoretic model for the Hodge theory within the framework of BRST formalism. In fact, such a proof 
is expected to provide a systematic theoretical basis for the existence of 
the quantum fields with negative kinetic terms ({\it but} with well-defined mass). These {\it exotic} fields are a set of
possible candidates for the dark matter (whose massless limit corresponds to dark energy). Finally, we wish to study the higher $p$-form ($p = 2, 3...$)
gauge theories up to $p = 5$ forms because these tower of quantum  
fields are relevant in the context of the {\it quantum} self-dual superstring theories that are consistently defined in the
$D = 10$ dimensions of spacetime.

The theoretical contents of our present endeavor are organized as follows. In Sec. 2, we show the existence of the first-class 
constraints for the {\it massless} Abelian 3-form theory in any arbitrary D-dimension of spacetime. We write down the generator
for the {\it classical}  gauge symmetry transformation in terms of {\it these} first-class constraints and the non-trivial canonical commutators are mathematically expressed.   Our Sec. 3  deals with the derivations of the Noether conserved current
and charge where we establish a connection between the first-class constraints and the Noether conserved charge.
Our Sec. 4 is devoted to a very brief discussion on the BRST approach to the D-dimensional massless Abelian 3-form theory
where we recapitulate the bare essentials of our earlier work [28]
and show that the standard Noether (anti-) BRST charges are non-nilpotent. 
In Sec. 5, we derive the off-shell nilpotent versions of the (anti-)BRST charges
from the standard non-nilpotent Noether (anti-)BRST charges.  
Our Sec. 6, contains the theoretical material on the physicality criteria w.r.t. the off-shell nilpotent versions 
of the (anti-)BRST charges where we show the appearance of the operator form of the first-class constraints. 
Finally, we make some concluding remarks in our Sec. 7 and point out a few future directions for further investigation(s).

In our Appendix A, we show the existence of the CF-type restrictions from the 
requirement of the absolute anticommutativity of the nilpotent 
(anti-)BRST charges. \\

{\it Convention and Notations}:
The background flat D-dimensional Minkowskian spacetime manifold is endowed with the metric tensor
 $\eta_{\mu\nu} =$ diag  $(+1, -1,  -1,...)$
so that the dot product between two non-null vectors $X^\mu$ and $Y^\mu$ in this space is: 
$X\cdot  Y = \eta_{\mu\nu}\, X^\mu \, Y^\nu \equiv 
 X_0 Y_0 - X_i Y_i$ where the Greek indices 
$\mu, \nu, \lambda,...  = 1, 2,..., D - 1$ correspond to the time and space directions and  the Latin indices $i, j, k,... = 1, 2, 3,...,D - 1$ stand for the space directions {\it only}. 
Einstein's summation convention is adopted throughout our present endeavor. The convention of the 
left-derivatives w.r.t. all the fermionic fields has been taken into consideration in the entire text of our present 
paper. We always denote the (anti-)BRST symmetry transformations by the symbols $s_{(a)b}$
and the corresponding (anti-)BRST charges are represented by $Q_{(a)b}$. The nilpotent [$s_{(a)b}^2 = 0$]
(anti-)BRST symmetry transformations are fermionic in nature and they anticommute with {\it all} the fermionic fields 
and commute with the bosonic fields of our theory.
We have also adopted the convention of the field derivatives as: $ (\partial\, B_{\mu\nu}/ \partial\,B_{\alpha\beta}) = \frac{1}{2 !}\, [\delta_\mu^\alpha \delta_\nu^\beta 
- \delta_\nu^\alpha \delta_\mu^\beta ], \; ({\partial A_{\mu\nu\lambda}}/{\partial A_{\rho\sigma\kappa}}) = 
\frac {1}{3!} \big [\delta_\mu ^\rho \,(\delta_\nu ^\sigma \, \delta _\lambda ^\kappa - \delta _\lambda ^\sigma\, \delta _\nu ^\kappa) + 
\delta_\nu ^\rho \, (\delta_\lambda ^\sigma \,\delta _\mu ^\kappa - \delta _\mu ^\sigma\, \delta _\lambda ^\kappa) + 
\delta_\lambda ^\rho \, (\delta_\mu^\sigma \,\delta _\nu ^\kappa - \delta _\nu ^\sigma\, \delta _\mu ^\kappa)  \big]$, etc., 
for the D-dimensional Abelian higher  $p$-form ($p = 2, 3,...$) fields.

\section{First-Class Constraints: Massless Abelian 3-Form Theory in Any Arbitrary Dimension of Spacetime}

We begin with the free Lagrangian density $({\cal L}_0)$ for the free Abelian 3-form 
$[A^{(3)} = \{(d\,x^\mu \wedge d\,x^\nu \wedge d\, x^\lambda)/3!\}\, A_{\mu\nu\lambda}]$ gauge theory 
in any arbitrary D-dimension  of the flat Minkowskian 
spacetime (which is described by {\it only} the kinetic term),  namely;
\begin{eqnarray}
{\cal L}_{0} = \frac {1}{24}\,H^{\mu \nu \lambda \zeta }\,H_{\mu \nu \lambda \zeta},
\end{eqnarray}
where $H^{(4)} = d\, A^{(3)} \equiv  [(d\,x^\mu \wedge d\,x^\nu \wedge d\, x^\lambda \wedge d\,x^\zeta)/4!]\,H_{\mu\nu\lambda\zeta}$
(with the explicit form: $H_{\mu \nu \lambda \zeta} = \partial_\mu\, A_{\nu\lambda\zeta} - \partial_\nu\, A_{\lambda \zeta \mu } 
 + \partial_\lambda\, A_{\zeta \mu \nu }  - \partial_\zeta \,A_{\mu \nu \lambda }$) defines  the field 
 strength tensor  $H_{\mu\nu\lambda\zeta}$ for the {\it totally}  antisymmetric gauge field $A_{\mu\nu\lambda}$. 
 Here the symbol: $d = d x^\mu\, \partial_\mu$ (with $d^2 = 0, \; \mu = 0, 1, 2, ... D-1$) stands for the exterior derivative of the 
 differential geometry [29-33].  
 The totally antisymmetric nature\footnote{The origin of the totally antisymmetric field-strength 
tensor $H_{\mu\nu\lambda \zeta}$ from $H^{(4)} = d A^{(3)}$ automatically demonstrates that: $A^{(3)} \longrightarrow A^{(3)} + d\, \Lambda^{(2)}$ is the
 underlying gauge symmetry of 
$H^{(4)}$ because of the nilpotency (i.e. $d^2 = 0$) of the exterior derivative. Here the 2-form $\Lambda^{(2)} = [(dx^\mu \wedge dx^\nu)/2!]\, \Lambda_{\mu\nu}$ defines the antisymmetric 
($\Lambda_{\mu\nu} = -\, \Lambda_{\nu\mu}$) tensor gauge transformation parameter. In explicit tensorial  language,  
we have the transformation: $A_{\mu\nu\lambda} \longrightarrow A_{\mu\nu\lambda}  + (\partial_\mu \Lambda _{\nu\lambda} 
+ \partial_\nu \Lambda _{\lambda\mu} + \partial_\lambda \Lambda _{\mu\nu})$ under which $H_{\mu\nu\lambda\zeta}$
remains invariant.} of $H_{\mu\nu\lambda\zeta}$ forces the spacetime  to be $D \geq 4$. 
 It can be explicitly checked  that
the starting Lagrangian density $({\cal L}_0)$ is {\it singular}. Hence, there are constraints on the theory. 
The conjugate momenta,  w.r.t. the gauge field  $A_{\mu\nu\lambda}$,  are: 
\begin{eqnarray}
\Pi^{\mu\nu\lambda} = \frac {\partial {\cal L}_0}{\partial (\partial_0 A_{\mu\nu\lambda})} \;\; \equiv \;\;\frac {1}{3}\, H^{0\mu\nu\lambda}
\quad \Longrightarrow \quad \Pi^{0ij}  = \frac {1}{3}\, H^{00ij} \approx 0.
\end{eqnarray}
In the above, the quantity $\Pi^{0ij} \approx 0$
is the primary constraint on the theory where the Dirac notation of the symbol for the idea of {\it weakly} zero has been taken into account. 
It is straightforward to note that the Euler-Lagrange (EL) equation of motion (EoM)
\begin{eqnarray}
\partial_\mu H^{\mu\nu\lambda\zeta} = 0 \quad \Longrightarrow \quad \partial _0 H^{00jk} + \partial_i H^{i0jk} = 0,
\end{eqnarray}
is true where the choices:  $\nu = 0, \;\lambda = j$ and $\zeta = k$ have been taken into consideration. 
It is clear from equation (3) that we have the time derivative (in the natural units: $\hbar = c = 1$, $\partial_0 = \partial_t$) 
on the primary constraint $\Pi^{0ij}  \approx 0$ [cf. Eq. (2)] as follows [34]
\begin{eqnarray}
\frac {\partial \Pi ^{0jk}}{\partial t}\,  = \frac {1}{3}\,\partial_i H^{0ijk} \approx 0 \quad \Longrightarrow 
\quad \frac {\partial \Pi ^{0jk}}{\partial t} \equiv \partial_i \Pi ^{ijk} \approx 0,
\end{eqnarray}
where $\Pi ^{ijk} = \frac {1}{3}\, H^{0ijk}$  is the expression for the {\it purely}  space components  of the conjugate momenta. 
 There are {\it no} further  constraints on the theory after the derivation of the 
secondary constraint (i.e. $\partial_i \Pi ^{ijk} \approx 0$) in Eq. (4). It is interesting to point out that {\it both}  
the primary as well as the secondary constraints are expressed in terms of  
the components  of the canonical conjugate momenta [defined in Eq. (2) w.r.t. the gauge field $A_{\mu\nu\lambda}$]. 
Hence, they  commute with each-other leading to their characterization  as the first-class constraints in the terminology 
of Dirac's prescription for the classification scheme of constraints (see, e.g. [11-13] for details). Thus, we note that we have:
$\Pi ^{0ij} \approx 0$ and $\partial_i \Pi ^{ijk} \approx 0$ as the primary and secondary constraints, respectively, on our theory.

One of the key signatures of a classical gauge theory is the existence of the first-class constraints on it  (see, e.g. [11-15] for details). 
The expression for the generator of the local, continuous, infinitesimal {\it classical} gauge symmetry transformations,
 in terms of the first-class constraints, can be written as (see, e.g. [16, 17.] for details):
\begin{eqnarray}
G = \int d^{D-1} x \, \Big[\Pi^{0ij} (\partial_0\, \Lambda_{ij})  + \Pi^{ij0}\, (\partial_i\, \Lambda_{j0}) 
+ \Pi^{j0i}\,( \partial_j\, \Lambda_{0i})  \nonumber\\
- (\partial _i \Pi^{ijk})  \, \Lambda _{jk}   - (\partial_j \Pi ^{jki}) \, \Lambda_{ki}   - (\partial_k \Pi ^{kij}) \,\Lambda_{ij} \Big].
\end{eqnarray} 
In the above expression, the totally antisymmetric nature of $\Pi ^{0ij}$ and $\Pi ^{ijk}$ has been taken into account
along with that of $\Lambda_{0i}$ and $\Lambda_{ij}$ where a set of local (i.e. spacetime dependent)
functions ($\Lambda_{\mu\nu} (x) \equiv \Lambda_{0i} (x), \Lambda_{ij} (x)$) are the antisymmetric ($\Lambda_{\mu\nu} = -\, \Lambda_{\nu\mu}$) tensor gauge 
symmetry transformation parameters. 
Using the Gauss divergence theorem, we can re-express the above generator, in {\it its}  more transparent and useful  form, as  
\begin{eqnarray}
G = \int d^{D-1} x \, \Big[\Pi^{0ij} \; (\partial_0\, \Lambda_{ij} + \partial_i\, \Lambda_{j0} + \partial_j\, \Lambda_{0i})  \nonumber\\
+  \Pi^{ijk} \; (\partial_i\, \Lambda_{jk} + \partial_j\, \Lambda_{ki} + \partial_k\, \Lambda_{ij}) \Big]. 
\end{eqnarray} 
 The above generator should be able to generate the infinitesimal and local 
gauge symmetry transformation: $\delta_g A_{\mu\nu\lambda}  = \partial _\mu \Lambda _{\nu\lambda} + \partial _\nu \Lambda _{\lambda\mu} + 
\partial _\lambda \Lambda _{\mu\nu}$ under which the starting Lagrangian density $({\cal L}_0)$ remains invariant 
(i.e. $\delta_g {\cal L}_0 = 0$). Using the following {\it non-trivial} equal-time canonical commutators  
\begin{eqnarray}
\big[A_{0ij} \, (\vec x, t), \, \Pi^{0kl}_{(A)}\, (\vec y, t)\big] &=& \frac{i}{2!}\, \big( \delta^k_i\, \delta^l_j -  \delta^l_i\, \delta^k_j \big)\, 
\delta^{(D-1)}\, (\vec x - \vec y), \nonumber\\
\big[A_{ijk} \, (\vec x, t), \, \Pi^{lmn}_{(A)}\, (\vec y, t)\big] &=& \frac{i}{3!}\, 
\big[\delta_i^l\, \big( \delta^m_j\, \delta^n_k -  \delta^n_j\, \delta^m_k \big)
+ \delta_i^m\, \big( \delta^n_j\, \delta^l_k -  \delta^l_j\, \delta^n_k \big) \nonumber\\
 &+& \delta_i^n\, \big( \delta^l_j\, \delta^m_k -  \delta^m_j\, \delta^l_k \big)
\big] \, \delta^{(D-1)}\, (\vec x - \vec y), 
\end{eqnarray}
it can be easily checked that we have the following 
\begin{eqnarray}
&&\delta_g\, A_{0ij}\,(\vec x, t) = -\, i\, \big[A_{0ij}\,(\vec x, t), G  \big] \equiv 
(\partial_0\, \Lambda_{ij} + \partial_i\, \Lambda_{j0} + \partial_j\, \Lambda_{0i}),   \nonumber\\
&&\delta_g\, A_{ijk}\,(\vec x, t) = -\, i\, \big[A_{ijk}\,(\vec x, t), G  \big] \equiv 
(\partial_i\, \Lambda_{jk} + \partial_j\, \Lambda_{ki} + \partial_k\, \Lambda_{ij}),
\end{eqnarray}
where $(A_{0ij}, \; A_{ijk})$ are the {\it independent} components of the totally antisymmetric  gauge field $A_{\mu\nu\lambda}$.
We would like to point out that {\it all} the rest of the other brackets [except (7)] of the theory are {\it trivially} zero. 
The above observation (8) establishes  that the first-class constraints generate the local, infinitesimal 
and continuous gauge symmetry transformations.

We end this section with the following concluding remarks. First, we note that the Lagrangian  density/Lagrangian of a 
gauge theory is always {\it singular} which implies that there are constraints on any arbitrary gauge theory. Second, for a gauge theory, the 
constrains are of the first-class variety in the terminology of Dirac's prescription for the classification scheme of constraints [11-15]. 
Finally, the local, 
infinitesimal and continuous gauge symmetry transformations owe their origin to the existence of the first-class constraints 
on the theory.   
Hence, the existence of the first-class constraints is a {\it decisive} signature of a theory to be an example of  gauge theory.
 Thus, the massless Abelian 3-form theory, described  by the Lagrangian density (1), is 
a simple example of a  gauge theory.

\section{Noether Conserved Charge and  First-Class Constraints of the  Theory  Under Consideration}

In this section, we establish a  connection between the Noether conserved charge 
 (corresponding to the infinitesimal gauge symmetry transformation)
and the first-class constraints of the theory  under consideration. In this context, first of all, 
we note that the Noether conserved  current,  for the massless Abelian  3-form gauge theory,  is 
\begin{eqnarray}
J^\mu  =  (\delta_g A_{\nu\lambda\xi})\, \frac {\partial {\cal L}_0}{\partial(\partial_\mu A_{\nu\lambda\xi})} \equiv  \frac {1}{3}\, H^{\mu\nu\lambda\zeta}\, (\partial_\nu\Lambda_{\lambda\zeta} + \partial_\lambda\Lambda_{\zeta\nu} 
+  \partial_\zeta\Lambda_{\nu\lambda}),
\end{eqnarray}
because of the observation that $\partial_\mu J^\mu = 0$ due to the EoM ($\partial_\mu H^{\mu\nu\lambda\zeta} = 0$)
and the totally antisymmetric property of $H_{\mu\nu\lambda\zeta}$ and the symmetric property of the two successive ordinary derivatives
on the gauge symmetry transformation parameter $(\Lambda_{\mu\nu})$. 
The above Noether conserved current leads to the definition of the  conserved charge:  
\begin{eqnarray}
Q = \int d^{D - 1} x\, J^0 = \int d^{D - 1} x \Big[\frac {1}{3}\, H^{0\nu\lambda\zeta}\, (\partial_\nu\Lambda_{\lambda\zeta} 
+ \partial_\lambda\Lambda_{\zeta\nu} +  \partial_\zeta\Lambda_{\nu\lambda}) \Big].
\end{eqnarray}
Since our theory is endowed with the constraints (e.g. $\Pi^{0ij} = \frac {1}{3}H^{00ij} \approx 0, 
\; \partial_i \Pi ^{ijk} =  \frac {1}{3}\; \partial_i H^{0ijk} \approx 0$), we have to be 
careful in expanding the r.h.s. of the above charge. In other words, 
we can {\it not} set the constraints strongly equal to zero. Thus,  we have 
\begin{eqnarray}
Q = \int d^{D - 1} x \Big[\frac {1}{3}\, H^{00ij}\, (\partial_0\Lambda_{ij} 
+ \partial_i\Lambda_{j0} +  \partial_j\Lambda_{0i}) \nonumber\\
 + \frac {1}{3}\, H^{0ijk}\, (\partial_i\Lambda_{jk} 
+ \partial_j\Lambda_{ki} +  \partial_k\Lambda_{ij}) \Big],
\end{eqnarray}
which is nothing but the {\it final} form of the generator [cf. Eq. (6)] for the gauge symmetry transformation
($\delta_g A_{\mu\nu\lambda} = \partial_\mu \Lambda_{\nu\lambda} + \partial_\nu \Lambda_{\lambda\mu} + \partial_\lambda \Lambda_{\mu\nu}$)
in the case of the {\it massless} Abelian 3-form gauge theory (with the identifications: 
$\Pi^{0ij} = \frac {1}{3}H^{00ij}, \; \Pi^{ijk} = \frac {1}{3}H^{0ijk}$).
It is clear that, in a subtle manner, the first-class constraints are hidden in (6)
as well as in the expression for the Noether conserved charge $Q$ corresponding to the Noether conserved current.

We conclude this short section with the following useful remarks. First, there exists a connection between the standard 
generator of the infinitesimal, local and continuous gauge symmetry transformation that is expressed in terms of the first-class
constraints and the Noether conserved charge (that is computed from the Noether conserved current) by using the 
celebrated Noether theorem. Second, in the computation of the conserved charge, the constraints should {\it not} be set equal to zero 
 strongly if they appear in the zeroth component of the Noether conserved current. 
Finally, a close and careful look at expressions for the standard generator 
[16, 17] for the gauge symmetry transformation and the conserved charge 
demonstrates that both are  one and the same.  
Thus, the Noether conserved charge is nothing but the generator for the infinitesimal, 
local and continuous gauge symmetry transformation for our D-dimensional massless Abelian 3-form theory.

\section{BRST Approach: Some Key Points}

For our paper to be self-contained, we recapitulate the bare essentials of our earlier work [28] and highlight
some of the issues that have {\it not} been pointed out strongly in [28].
This section is divided into two subsections. In Subsec. 4.1, we write down the coupled (but equivalent) 
Lagrangian densities and show their (anti-)BRST invariance with a bit of emphasis on the CF-type restrictions. 
The subject matter of Subsec. 4.2 is concerned  with the derivations of the standard Noether conserved currents and charges. 
We lay emphasis on the non-nilpotency property of the standard Noether (anti-) BRST charges that are the generators for the (anti-)BRST symmetries.

\subsection{Coupled Lagrangian Densities: (Anti-)BRST Invariance}

We begin with the description of the coupled\footnote{ The (anti-)BRST invariant Lagrangian densities are coupled because
of the existence of the non-trivial CF-type restrictions [cf. Eq. (20) below] which connect some of the auxiliary as well as the
basic fields of {\it both} the Lagrangian densities in a very specific fashion. 
Hence, some of these connected fields are {\it not} independent. 
These Lagrangian densities are also {\it equivalent} from the point of view of the symmetry transformations as both of them respect the
(anti-)BRST symmetry transformations provided the validity of the (anti-)BRST invariant CF-type restrictions are taken into account
(see, e.g. [28] for details).} (but equivalent) (anti-) BRST invariant Lagrangian densities that 
incorporate into themselves the gauge-fixing and FP-ghost terms. These Lagrangian densities for the massless Abelian 
3-form theory have been derived in our earlier work [28] which are nothing but the generalizations of the original 
Lagrangian density $[{\cal L}_0 = (1/24)\, H^{\mu\nu\lambda\xi}\, H_{\mu\nu\lambda\xi}]$. 
The explicit form of the coupled (but equivalent) (anti-) BRST invariant Lagrangian densities  
are as follows (see, e.g. [28])
\begin{eqnarray}
{\cal L}_{\bar B} &=& \frac{1}{24} H^{\mu\nu\eta\xi} H_{\mu\nu\eta\xi}
-  {\bar B}^{\mu\nu} \Bigl ( \partial^\eta A_{\eta\mu\nu} - \frac{1}{2} 
[\partial_\mu \phi_\nu - \partial_\nu \phi_\mu] \Bigr )
 - \frac{1}{2} {\bar B}_{\mu\nu} {\bar B}^{\mu\nu}\nonumber\\
 &+& (\partial_\mu \bar C_{\nu\eta} + \partial_\nu \bar C_{\eta\mu} 
+ \partial_\eta \bar C_{\mu\nu}) (\partial^\mu C^{\nu\eta}) + (\partial \cdot \phi)\,  B_1 - \frac{1}{2} B_1^2 - B B_2 \nonumber\\
&-& (\partial_\mu \bar \beta_\nu - \partial_\nu \bar \beta_\mu) (\partial^\mu \beta^\nu) 
- (\partial_\mu \bar C^{\mu\nu} + \partial^\nu \bar C_1) F_\nu + (\partial_\mu  C^{\mu\nu} 
- \partial^\nu C_1) \bar f_\nu \nonumber\\
&+& \partial_\mu \bar C_2 \partial^\mu C_2 + (\partial \cdot \beta) B_2 - (\partial \cdot \bar \beta) B,
\end{eqnarray}
\begin{eqnarray}
{\cal L}_B &=& \frac{1}{24} H^{\mu\nu\eta\xi} H_{\mu\nu\eta\xi}
+  B^{\mu\nu} \Bigl ( \partial^\eta A_{\eta\mu\nu} + \frac{1}{2} 
[\partial_\mu \phi_\nu - \partial_\nu \phi_\mu] \Bigr )
 - \frac{1}{2} B_{\mu\nu}  B^{\mu\nu}\nonumber\\
&+& (\partial_\mu \bar C_{\nu\eta} + \partial_\nu \bar C_{\eta\mu} 
+ \partial_\eta \bar C_{\mu\nu})(\partial^\mu C^{\nu\eta}) + (\partial \cdot \phi)\, B_1 - \frac{1}{2} B_1^2 - B B_2\nonumber\\
&-& (\partial_\mu \bar \beta_\nu - \partial_\nu \bar \beta_\mu) (\partial^\mu \beta^\nu) 
+ (\partial_\mu \bar C^{\mu\nu} + \partial^\nu \bar C_1) f_\nu  -  (\partial_\mu  C^{\mu\nu} 
- \partial^\nu C_1) \bar F_\nu \nonumber\\
&+& \partial_\mu \bar C_2 \partial^\mu C_2 + (\partial \cdot \beta) B_2 - (\partial \cdot \bar \beta) B,
\end{eqnarray}
where, as pointed out earlier, the field-strength tensor $H_{\mu\nu\lambda\xi}$ is derived from:
$H^{(4)} = d\, A^{(3)}$. 
The gauge-fixing term owes its origin to the co-exterior derivatives 
$\delta = \mp * d * $ because we note that the following 2-form 
 \begin{eqnarray}
\delta\, A^{(3)} = -\, * d * A^{(3)} = 
\Big(\frac {d\, x^\mu \wedge d\, x^\nu }{2!}  \Big) \Big(\partial^\lambda\, A_{\lambda\mu\nu}\Big),
\end{eqnarray}
defines the gauge-fixing term (i.e. $\partial^\lambda\, A_{\lambda\mu\nu}$) for the 3-form gauge field $A_{\lambda\mu\nu}$
where we have taken a minus sign on the r.h.s. for the sake of brevity which is true for the {\it even} 
dimensional flat Minkowskian spacetime manifold. 
Here, there is a room for its generalization because we can add/subtract a 2-form 
$\Phi^{(2)} = d\, \Phi^{(1)} \equiv [(d x^{\mu} \wedge d x^{\nu})\, 2 !]\, 
(\partial_\mu\, \phi_\nu - \partial_\nu\, \phi_\mu)$ where the 1-form $\Phi^{(1)} = d x^\mu \, \phi_\mu$
defines a vector field $\phi_\mu$. This is why, we have the gauge-fixing terms in 
${\cal L}_{ \bar B}$ and ${\cal L}_{B}$ as: 
$ (\partial^\lambda\, A_{\lambda\mu\nu} - \frac{1}{2} 
[\partial_\mu \phi_\nu - \partial_\nu \phi_\mu])$ and 
$(\partial^\lambda\, A_{\lambda\mu\nu} + \frac{1}{2} 
[\partial_\mu \phi_\nu - \partial_\nu \phi_\mu])$, respectively. 
The fermionic (anti-)ghost fields $(\bar C_{\mu\nu})\, C_{\mu\nu}$ are the generalizations of the 
classical level gauge symmetry transformation parameter $\Lambda_{\mu\nu}$ and they carry the ghost numbers 
$(-1)+1$, respectively.  The bosonic (anti-)ghost fields $(\bar \beta_\mu)\beta_\mu$ are endowed with the 
ghost numbers $(-2)\, +2$, respectively. On the other hand, we have the fermionic 
(anti-)ghost fields $(\bar C_2)\, C_2$ that carry the ghost numbers (-3)+3, respectively. 
Furthermore, there are bosonic auxiliary fields ($B, B_1, B_2$) and the fermionic auxiliary fields
($f_\mu, \bar f_\mu, F_\mu, \bar F_\mu$) along with a bosonic vector field $(\phi_\mu)$ in our theory.
 It can be checked  that the auxiliary   fields $B$ and $B_2$ carry the ghost number $(+2)$ and $(-2)$, respectively, 
because we observe  that: $B = (\partial\cdot\beta)$ and $B_2 = -\,(\partial\cdot\bar\beta)$. 
However, the auxiliary field $B_1 = (\partial\cdot\phi)$ carries the ghost number equal to zero. 
The fermionic auxiliary fields in pairs $(f_\mu, \, F_\mu)$ and $(\bar f_\mu,\, \bar F_\mu )$ 
are endowed with the ghost numbers $(+1)$ and $(-\, 1)$, respectively. Similarly, the (anti-)ghost fermionic fields 
$(\bar C_1)C_1$ carry the ghost numbers $(-1)+1$, respectively. 
The Nakanishi-Lautrup type auxiliary fields $(B_{\mu\nu}, \bar B_{\mu\nu})$ are invoked to 
linearize the gauge-fixing terms in ${\cal L}_B$ and  ${\cal L}_{\bar B}$, respectively, and they are connected to each-other by 
the (anti-)BRST invariant CF-type restriction
(i.e. $B_{\mu\nu} + \bar B_{\mu\nu} = \partial_\mu\, \phi_\nu - \partial_\nu\phi_\mu$).

The generalizations of the {\it classical} gauge symmetry transformation to their {\it quantum} 
counterparts (anti-)BRST symmetry transformations [$s_{(a)b}$] are as follows (see, e.g. [28])
\begin{eqnarray}
&&s_{ab} A_{\mu\nu\eta} = \partial_\mu \bar C_{\nu\eta} + \partial_\nu \bar C_{\eta\mu}
+ \partial_\eta \bar C_{\mu\nu}, \;\;  s_{ab} \bar C_{\mu\nu} = \partial_\mu \bar \beta_\nu
- \partial_\nu \bar \beta_\mu, \;\; s_{ab}  C_{\mu\nu} = \bar B_{\mu\nu}, \nonumber\\
&&s_{ab} B_{\mu\nu} = \partial_\mu \bar f_\nu - \partial_\nu \bar f_\mu, \;\;
s_{ab}  \beta_\mu =  F_\mu, \;\;
s_{ab} \bar \beta_\mu = \partial_\mu \bar C_2, \;\; s_{ab} \bar F_\mu = - \partial_\mu B_2, \nonumber\\
&&s_{ab} C_2 = B, \;\; s_{ab} f_\mu = - \partial_\mu B_1, \;\; s_{ab} C_1 = - B_1, \;\; s_{ab} \bar C_1 = - B_2, 
\quad s_{ab} \phi_\mu = \bar f_\mu,\nonumber\\
&&s_{ab}\; \Bigl [ \bar C_2, \;\bar f_\mu, \;F_\mu,\; B,\; B_1, \;B_2, \;\bar B_{\mu\nu}, \;H_{\mu\nu\eta\kappa}
\;\Bigl ] \;= \;0,
\end{eqnarray}
\begin{eqnarray}
&&s_b A_{\mu\nu\eta} = \partial_\mu C_{\nu\eta} + \partial_\nu C_{\eta\mu}
+ \partial_\eta C_{\mu\nu}, \;\; s_b C_{\mu\nu} = \partial_\mu \beta_\nu
- \partial_\nu \beta_\mu, \;\; s_b \bar C_{\mu\nu} = B_{\mu\nu}, \nonumber\\
&&s_b \bar B_{\mu\nu} = \partial_\mu f_\nu - \partial_\nu f_\mu, \;\;
s_b \bar \beta_\mu = \bar F_\mu, \;\;
s_b \beta_\mu = \partial_\mu C_2, \;\; s_b F_\mu = - \partial_\mu B, \nonumber\\
&&s_b {\bar C}_2 = B_2, \;\; s_b C_1 = - B, \;\; s_b \bar C_1 = B_1, \;\; s_b \phi_\mu = f_\mu, 
\;\;s_b \bar f_\mu = \partial_\mu B_1,\nonumber\\
&&s_b \; \Bigl [ C_2,\; f_\mu,\; {\bar F}_\mu,\; B,\; B_1,\; B_2,\; B_{\mu\nu}, \;H_{\mu\nu\eta\kappa} \;\Bigl ]
\; = \;0,
\end{eqnarray}
which are found to be off-shell nilpotent [$s_{(a)b}^2 = 0$] of order two. Hence, these {\it quantum} transformations 
are fermionic in nature and they transform the bosonic fields into the fermionic fields and vice-versa. 
It is straightforward to 
check that the Lagrangian densities ${\cal L}_B$ and  ${\cal L}_{\bar B}$ transform, under the (anti-) BRST
symmetry transformations, to the total spacetime derivatives as:
\begin{eqnarray}
s_{ab} {\cal L}_{\bar B} &=& \partial_\mu \Bigl [- (\partial^\mu {\bar C}^{\nu\eta} + \partial^\nu {\bar C}^{\eta\mu}
+ \partial^\eta {\bar C}^{\mu\nu}) \bar B_{\nu\eta} + {\bar B}^{\mu\nu} {\bar f}_\nu - (\partial^\mu {\bar \beta}^\nu -
\partial^\nu {\bar \beta}^\mu) F_\nu \nonumber\\
&+& B_1 {\bar f}^\mu + B_2 F^\mu - B \partial^\mu {\bar C}_2 \Bigr ].
\end{eqnarray}
\begin{eqnarray}
s_b {\cal L}_ B &=&  \partial_\mu \Bigl [ (\partial^\mu C^{\nu\eta} + \partial^\nu C^{\eta\mu}
+ \partial^\eta C^{\mu\nu})  B_{\nu\eta} + B^{\mu\nu} f_\nu 
- (\partial^\mu \beta^\nu - \partial^\nu \beta^\mu) \bar F_\nu \nonumber\\
&+& B_1 f^\mu - B \bar F^\mu + B_2 \partial^\mu C_2 \Bigr ].
\end{eqnarray}
Hence, the action integrals $S_1 = \int d^D x\,{\cal L}_{\bar B}$ and 
$S_2 = \int d^D x\,{\cal L}_{B}$ remain  invariant (i.e $s_{ab}\, S_1 = 0, \, s_b\, S_2 = 0$) under
the (anti-)BRST symmetrry transformations, respectively.

We  dwell a bit on the anticommutativity property of the (anti-)BRST symmetry transformations
and lay emphasis on the CF-type restrictions. 
It turns out that the absolute anticommutativity property (i.e. $\{s_b, s_{ab}\} = 0$) is satisfied for {\it all} the fields 
of our theory {\it except} fields $A_{\mu\nu\lambda}, C_{\mu\nu}$ and $\bar C_{\mu\nu}$ because we observe the following: 
 \begin{eqnarray}
&& \{s_b, s_{ab}\}\, A_{\mu\nu\lambda} = \partial_\mu (B_{\nu\lambda} +  \bar B_{\nu\lambda}) + \partial_\nu (B_{\lambda\mu}
 +  \bar B_{\lambda\mu}) + \partial_\lambda (B_{\mu\nu} +  \bar B_{\mu\nu}), \nonumber\\
 && \{s_b, s_{ab}\}\, C_{\mu\nu} = \partial_\mu (f_\nu +  F_\nu) - \partial_\nu (f_\mu + F_\mu), \nonumber\\
  && \{s_b, s_{ab}\}\,\bar C_{\mu\nu} = \partial_\mu (\bar f_\nu +  \bar F_\nu) - \partial_\nu (\bar f_\mu + \bar F_\mu).
\end{eqnarray}  
However, for the above fields, we have the validity of the absolute  anticommutativity property (i.e.  $\{s_b, s_{ab}\} = 0$) of
the (anti-)BRST transformations provided we use the following CF-type restrictions
which have been derived by using the superfield approach to BRST formalism [23, 24]:
\begin{eqnarray}
B_{\mu\nu} + \bar B_{\mu\nu} = \partial_\mu \phi_\nu  - \partial_\nu \phi_\mu, \qquad 
f_\mu + F_\mu  = \partial_\mu C_1, \qquad \bar f_\mu + \bar F_\mu  = \partial_\mu \bar C_1.
\end{eqnarray} 
Using the (anti-)BRST symmetry transformations [cf. Eqs. (15), (16)], it can be checked that the CF-type restrictions 
are (anti-)BRST invariant quantities, namely; 
\begin{eqnarray}
&&s_{(a)b}[B_{\mu\nu} + \bar B_{\mu\nu} - (\partial_\mu \phi_\nu  - \partial_\nu \phi_\mu)] = 0, \quad 
s_{(a)b}[f_\mu + F_\mu  - \partial_\mu C_1] = 0,\nonumber\\
&& s_{(a)b}[\bar f_\mu + \bar F_\mu  - \partial_\mu \bar C_1] = 0.
\end{eqnarray} 
Hence, these restrictions on our theory are {\it physical} and they are connected with the geometrical objects called gerbes [28, 35].
The coupled Lagrangian densities lead to the following EL-EoMs w.r.t. the fields: $B_{\mu\nu}, \bar B_{\mu\nu}, C_\mu, \bar C_\mu$
respectively: 
\begin{eqnarray}
&&B_{\mu\nu} = \partial^\lambda \, A_{\lambda\mu\nu} + \frac{1}{2}\, (\partial_\mu\, \phi_\nu - \partial_\nu\, \phi_\mu), \;\; 
\bar B_{\mu\nu} = -\, \partial^\lambda \, A_{\lambda\mu\nu} + \frac{1}{2}\, (\partial_\mu\, \phi_\nu - \partial_\nu\, \phi_\mu), \nonumber\\
&&\partial_\mu\, (\partial^\mu\, \bar C^{\nu\lambda} + \partial^\nu\, \bar C^{\lambda\mu} + \partial^\lambda\, \bar C^{\mu\nu})
+ \frac{1}{2}\, (\partial^\nu\, \bar{F}^\lambda - \partial^\lambda \, \bar F^\nu) = 0, \nonumber\\
&&\partial_\mu\, (\partial^\mu\, \bar C^{\nu\lambda} + \partial^\nu\, \bar C^{\lambda\mu} + \partial^\lambda\, \bar C^{\mu\nu})
- \frac{1}{2}\, (\partial^\nu\, \bar{f}^\lambda - \partial^\lambda \, \bar f^\nu) = 0, \nonumber\\
&&\partial_\mu\, (\partial^\mu\,  C^{\nu\lambda} + \partial^\nu\,  C^{\lambda\mu} + \partial^\lambda\,  C^{\mu\nu})
+ \frac{1}{2}\, (\partial^\nu\, {f}^\lambda - \partial^\lambda \,  f^\nu) = 0, \nonumber\\
&&\partial_\mu\, (\partial^\mu\,  C^{\nu\lambda} + \partial^\nu\,  C^{\lambda\mu} + \partial^\lambda\,  C^{\mu\nu})
- \frac{1}{2}\, (\partial^\nu\, {F}^\lambda - \partial^\lambda \,  F^\nu) = 0. 
\end{eqnarray}
It is clear, from the above, that we obtain the following: 
\begin{eqnarray}
&&B_{\mu\nu} + \bar B_{\mu\nu} = (\partial_\mu\, \phi_\nu - \partial_\nu\, \phi_\mu), \quad 
\partial_\mu\, (\bar f_\nu + \bar F_\nu) - \partial_\nu\,(\bar f_\mu + \bar F_\mu) = 0, \nonumber\\
&&\partial_\mu\, (f_\nu + F_\nu) - \partial_\nu\,(f_\mu + F_\mu) = 0.
\end{eqnarray}
Thus, we have already derived  the bosonic  CF-type restriction: 
$B_{\mu\nu} + \bar B_{\mu\nu} = (\partial_\mu\, \phi_\nu - \partial_\nu\, \phi_\mu)$ and 
we argue that the other two equations in (23) lead to the derivations of the fermionic CF-type restrictions: 
$f_\mu + F_\mu = \partial_\mu\, C_1$ and $\bar f_\mu + \bar F_\mu = \partial_\mu\, \bar C_1$. 
It is clear,  from the fermionic relationships  in (23), that the ghost numbers can be conserved iff we take into 
account the following non-trivial\footnote{There is a trivial solution where we can take, in a straightforward fashion, the combinations:
$f_\mu + F_\mu = 0$ and $\bar f_\mu + \bar F_\mu =0$. However, these combinations are {\it not} (anti-)BRST invariant. Hence, they can not
be a set of physical restrictions on our theory.} 
combinations of the pairs $(\bar f_\mu, \, \bar F_\mu)$ and $(f_\mu, \,  F_\mu)$ of the fermionic auxiliary fields and the 
(anti-)ghost fields $(\bar C_1)C_1$, namely; 
\begin{eqnarray}
f_\mu + F_\mu =  \pm \partial_\mu\, C_1, \qquad  \bar f_\mu + \bar F_\mu = \pm \partial_\mu\, \bar C_1, 
\end{eqnarray}
to make the r.h.s. of (23) equal to zero absolutely. However, the requirement of the (anti-)BRST invariance of the
CF-type restrictions ensure that {\it only} the positive signs on the r.h.s. of (24) are permitted. 
We would like to lay emphasis  that the r.h.s. of (24) can {\it not} be anything other than the derivatives  
on the (anti-)ghost fields $(\bar C_1)C_1$. In other words, {\it no}  other independent (anti-)ghost 
fields of our theory are permitted on the r.h.s. of (24).

We end this subsection with the concluding remarks that the coupled Lagrangian densities can yield the CF-type
restrictions from the EL-EoM. In other words, the {\it latter} are responsible for the existence of the coupled 
Lagrangian densities ${\cal L}_{B}$ and ${\cal L}_{\bar B}$. These Lagrangian densities are {\it equivalent} 
from the point of view of the symmetry considerations as both of these respect the (anti-)BRST symmetry transformations 
on the submanifold of the quantum Hilbert space of fields where all the three (anti-)BRST invariant CF-type restrictions
(i.e. $B_{\mu\nu} + \bar B_{\mu\nu} = \partial_\mu\, \phi_\nu - \partial_\nu\, \phi_\mu, 
\,\;  f_\mu + F_\mu = \partial_\mu\, C_1, \; \bar f_\mu + \bar F_\mu = \partial_\mu\, \bar C_1$) 
are satisfied (see, e.g. [28] for details).  \\

\subsection{Conserved Currents and Charges: Key Issues}

According to the celebrated Noether theorem, the symmetry invariance of the action integral leads to the 
derivations of the conserved Noether currents and corresponding charges depending on the number of 
{\it continuous} symmetry transformations that are respected by  the theory.
The continuous,  infinitesimal and 
off-shell nilpotent (anti-)BRST symmetry transformations [cf. Eqs. (15), (16)] lead to the following expressions for the Noether  conserved 
(anti-)BRST currents [28]:   
\begin{eqnarray}
J^\mu_{ab} &=& H^{\mu\nu\eta\xi} (\partial_\nu \bar C_{\eta\xi}) 
+ (\partial^\mu \bar C^{\nu\eta} + \partial^\nu \bar C^{\eta\mu}
+ \partial^\eta \bar C^{\mu\nu}) \; \bar B_{\nu\eta} + \bar B^{\mu\nu} \bar f_\nu 
+ B_1 \bar f^\mu \nonumber\\
&-& B \partial^\mu \bar C_2 + B_2 F^\mu - (\partial^\mu  \beta^\nu 
- \partial^\nu  \beta^\mu) (\partial_\nu \bar C_2)
-  (\partial^\mu \bar \beta^\nu - \partial^\nu \bar \beta^\mu) F_\nu \nonumber\\
&+& (\partial^\mu  C^{\nu\eta} + \partial^\nu  C^{\eta\mu}
+ \partial^\eta  C^{\mu\nu}) \; (\partial_\nu \bar \beta_\eta - \partial_\eta \bar \beta_\nu),\nonumber\\
J^\mu_{b} &=& H^{\mu\nu\eta\xi} (\partial_\nu C_{\eta\xi}) + (\partial^\mu C^{\nu\eta} 
+ \partial^\nu C^{\eta\mu} + \partial^\eta C^{\mu\nu}) \; B_{\nu\eta} + B^{\mu\nu} f_\nu 
+ B_1 f^\mu \nonumber\\
&+& B_2 \partial^\mu C_2 - B \bar F^\mu - (\partial^\mu \bar \beta^\nu 
- \partial^\nu \bar \beta^\mu) (\partial_\nu C_2) -  (\partial^\mu \beta^\nu 
- \partial^\nu \beta^\mu) \bar F_\nu \nonumber\\
& -& (\partial^\mu \bar C^{\nu\eta} + \partial^\nu \bar C^{\eta\mu}
+ \partial^\eta \bar C^{\mu\nu}) \; (\partial_\nu \beta_\eta - \partial_\eta \beta_\nu). 
\end{eqnarray}
The conserved (anti-)BRST charges (i.e. $Q_{ab}  = \int d^{D - 1} x \,{J}^0_{ab}$ 
and $Q_{b}  = \int d^{D - 1} x \,{J}^0_{b}$) from the above currents are as follows: 
\begin{eqnarray}
Q_{ab} &=& {\displaystyle \int} d^3 x \Bigl [ H^{0ijk} (\partial_i \bar C_{jk}) - 
 (\partial^0 \bar C^{ij}
+ \partial^i \bar C^{j 0} + \partial^j \bar C^{0i}) \bar B_{ij} + 
B_1 \bar f^0 \nonumber\\
&-& (\partial^0  \beta^i - \partial^i \beta^0) \partial_i \bar C_2 - 
(\partial^0  \bar\beta^i - \partial^i  \bar \beta^0) F_i
+ \bar B^{0i} \bar f_i - B \partial^0 {\bar C_2} \nonumber\\
&+&  (\partial^0  C^{ij} + \partial^i  C^{j 0}
+ \partial^j  C^{0 i}) (\partial_i \bar \beta_j - 
\partial_j \bar \beta_i) + B_2  F^0
\Bigr ],
\end{eqnarray}
\begin{eqnarray}
Q_{b} &=& {\displaystyle \int} d^3 x \Bigl [ H^{0ijk} (\partial_i C_{jk}) +  
(\partial^0 C^{ij} + \partial^i C^{j 0}
+ \partial^j C^{0 i}) B_{ij} + B_1 f^0 \nonumber\\
&-& (\partial^0 \bar \beta^i - \partial^i \bar \beta^0) \partial_i C_2 - 
(\partial^0  \beta^i - \partial^i  \beta^0) \bar F_i
+ B^{0i} f_i + B_2 \partial^0 C_2 \nonumber\\
&-&  (\partial^0 \bar C^{ij} + \partial^i \bar C^{j 0}
+ \partial^j \bar C^{0 i}) (\partial_i \beta_j - \partial_j \beta_i) - 
B \bar F^0
\Bigr ],
\end{eqnarray}
These conserved (anti-)BRST charges are the generators for the infinitesimal, continuous and off-shell nilpotent (anti-)BRST 
symmetry transformations (15) and (16) because the nilpotent (anti-)BRST symmetries for 
any generic  field $(\Phi)$ of the coupled (anti-)BRST invariant  Lagrangian densities 
${\cal L}_B$ and ${\cal L}_{\bar B}$ can be written as follows: 
\begin{eqnarray}
s_r \Phi  = -\,i\, [\Phi, \, Q_r]_{\pm}, \qquad \quad  r = ab, b,
\end{eqnarray}
where the superscripts $(\pm)$ signs  on the square bracket correspond to the  (anti)commutator for the generic field $\Phi$ being 
fermionic/bosonic in nature.

The relationship (28) is very general and it is applicable to any continuous symmetry transformation and its generator 
as the conserved Noether charge. For instance, it is obvious that: $s_b\, Q_b = -\, i\, \{Q_b, \, Q_b \} \equiv -\, 2\, i\, Q_b^2$ 
and $s_{ab} \, Q_{ab} = -\, i\, \{Q_{ab}, \, Q_{ab} \} \equiv -\, 2\, i\, Q_{ab}^2$.
In view of these relationships, it can be checked that we have the following 
\begin{eqnarray}
s_{ab} Q_{ab}  &=& \int d^{D- 1}x\, \Big[-\, (\partial^0 F^i - \partial^i F^0)\, \partial_i \bar C_2 \nonumber\\
 &+& (\partial^0 \bar B^{ij} + \partial^i \bar B^{j0}  
 + \partial^j \bar B^{0i})\, (\partial_i\bar \beta_j - \partial_j \bar \beta_i) \Big] \neq 0,
\end{eqnarray}
\begin{eqnarray}
s_{b} Q_{b}  &=& 
  \int d^{D- 1}x\, \Big[-\, (\partial^0 \bar F^i - \partial^i \bar F^0)\, \partial_i C_2 \nonumber\\
 &-& (\partial^0 B^{ij} + \partial^i B^{j0}  
 + \partial^j  B^{0i})\, (\partial_i \beta_j - \partial_j  \beta_i) \Big] \neq 0,
\end{eqnarray}
which demonstrate that the standard Noether conserved (anti-)BRST charges 
are {\it not} off-shell nilpotent (i.e. $Q_{(a)b}^2 \neq 0$) of order two. 
This happens due to the existence of the non-trivial CF-type restrictions (20) on our theory. 
In the above equations (29) and (30), the l.h.s. of each equation has been computed by the direct 
applications of the (anti-)BRST transformations 
(15) and (16) on the conserved (anti-)BRST charges (26) and (27).

We wrap-up this subsection with the following remarks. First of all, we have noted that the standard Noether (anti-)BRST
charges are computed from the infinitesimal, continuous and off-shell nilpotent (anti-)BRST symmetry transformations. However, 
these charges are found to be conserved {\it but} not off-shell nilpotent of order two. Second, one has the freedom 
to use the appropriate EL-EoMs and the Gauss divergence theorem to convert the non-nilpotent standard versions 
of the (anti-)BRST charges. Third, it turns out that the standard Noether theorem always leads to the derivations of the
non-nilpotent (anti-) BRST charges for the systems that are endowed with the non-trivial CF-type restrictions [22]. Finally, the 
BRST-quantized systems, with {\it trivial}  CF-type of restriction(s), lead to the existence of the conserved and off-shell 
nilpotent versions of the (anti-)BRST charges due to the standard Noether theorem. In this context, mention  can be made of the BRST-quantized 
theory of the D-dimensional  Abelian 1-form gauge theory. \\

\section{Nilpotent (Anti-)BRST Charges: Explicit Derivation from the Non-Nilpotent Noether Charges}

We follow our proposal [22] to obtain, in a systematic manner, the off-shell nilpotent versions of the (anti-)BRST 
charges [$Q_{(a)b}^{(1)}$] from the non-nilpotent versions of the (anti-)BRST charges (26) and (27) which have been derived by exploiting 
the standard theoretical techniques of Noether's theorem. In this context, first of all, we note that,  in the expression 
for the BRST charge (27), we can do the following 
\begin{eqnarray}
\int d^{D- 1}x \, H^{0ijk}\, (\partial_i C_{jk}) \equiv -\, \int d^{D- 1}x \, (\partial_i H^{0ijk})\, C_{jk}, 
\end{eqnarray}
where we have dropped a total space derivative term due to Gauss's divergence theorem because all the physical fields 
vanish off as $x \longrightarrow \pm\, \infty$. At this stage, we use the following EL-EoM that is derived from the Lagrangian 
density ${\cal L}_B$, namely; 
\begin{eqnarray}
\partial_\mu H^{\mu\nu\lambda\xi} + (\partial^\nu B^{\lambda\xi} + \partial^\lambda B^{\xi\mu} +  \partial^\xi B^{\mu\lambda}) = 0.  
\end{eqnarray}
Taking $\nu = 0, \lambda = j$ and $\xi = k$, we obtain the following 
\begin{eqnarray}
\partial_i H^{0ijk} = (\partial^0 B^{jk} + \partial^j B^{k0} +  \partial^k B^{0j}),   
\end{eqnarray}
where we have used the totally antisymmetric property of the field-strength tensor $H_{\mu\nu\lambda\xi}$. The 
substitution of (33) into the r.h.s. of (31) yields to the following: 
\begin{eqnarray}
-\, \int d^{D- 1}x \, (\partial^0 B^{ij} + \partial^i B^{j0} +  \partial^j B^{0i})\, C_{ij}.   
\end{eqnarray}
This term will be present in the off-shell nilpotent version of the BRST charge $Q_b^{(1)}$. An application of the BRST 
symmetry transformation (16) on (34) produces the following: 
\begin{eqnarray}
-\, \int d^{D- 1}x \, (\partial^0 B^{ij} + \partial^i B^{j0} +  \partial^j B^{0i})\, (\partial_i\beta_j - \partial_j \beta_i).   
\end{eqnarray}
This term should cancel out from an appropriate term of $Q_b$ [cf. Eq. (27)] when the BRST transformation is applied on it. 
For this purpose, we modify the following appropriate term of $Q_b$:
\begin{eqnarray}
&&-\, (\partial^0 C^{ij} + \partial^i C^{j0} +  \partial^j C^{0i})\, (\partial_i\beta_j - \partial_j \beta_i) \nonumber\\
&&=  
-\,2\,  (\partial^0 C^{ij} + \partial^i C^{j0} +  \partial^j C^{0i})\, (\partial_i\beta_j - \partial_j \beta_i)\nonumber\\
&&+  (\partial^0 C^{ij} + \partial^i C^{j0} +  \partial^j C^{0i})\, (\partial_i\beta_j - \partial_j \beta_i).  
\end{eqnarray}
It is clear that the application of $s_b$ on the {\it second} term of the r.h.s.  of (36) cancels (35). 
Thus, this term will be present in the expression for the off-shell nilpotent version of the BRST charge 
$Q_b^{(1)}$. In other words, so far, we have obtained {\it two} terms of $Q_b^{(1)}$ as: 
\begin{eqnarray}
\int d^{D- 1}x\,  \Big[(\partial^0 C^{ij} + \partial^i C^{j0} +  \partial^j C^{0i})\, (\partial_i\beta_j - \partial_j \beta_i) \nonumber\\
-\,  (\partial^0 B^{ij} + \partial^i B^{j0} +  \partial^j B^{0i})\, C_{ij}.
 \end{eqnarray}
It is straightforward to note that if we apply the BRST symmetry transformations (16) on (37), 
it will yield zero. In what follows, to obtain the explicit expression for $Q_b^{(1)}$, we shall exploit the theoretical
potential and power of 
(i) the Gauss divergence theorem, (ii) the BRST symmetry transformations, and (iii) the appropriate equations of motion
from ${\cal L}_B$.

At this juncture, we focus on the {\it first} term on the r.h.s. of (36). Because of the antisymmetry property 
in $i$ and $j$, we can write this term, taking into account the integral, as 
\begin{eqnarray}
&&-\,  4\,\int d^{D- 1}x \, (\partial^0 \bar C^{ij} + \partial^i \bar C^{j0} +  \partial^j \bar C^{0i})\, \partial_i\beta_j \nonumber\\
&&\; \equiv \; -\,  4\, \int d^{D- 1}x\, \partial_i\,  (\partial^0 \bar C^{ij} 
+ \partial^i \bar C^{j0} +  \partial^j \bar C^{0i})\, \beta_j
\end{eqnarray}
where we have applied the Gauss divergence theorem and dropped a total space derivative term. 
The following EL-EoM emerging out from ${\cal L}_B$, namely; 
\begin{eqnarray}
\partial_\mu\, (\partial^\mu\, \bar C^{\nu\lambda} + \partial^\nu\, \bar C^{\lambda\mu} + \partial^\lambda\, \bar C^{\mu\nu})
+ \frac{1}{2}\, (\partial^\nu\, \bar{F}^\lambda - \partial^\lambda \, \bar F^\nu) = 0,
\end{eqnarray}
with the choices: $\nu = 0, \, \lambda = j$ lead us to obtain the following:
\begin{eqnarray}
4\, \partial_i\,  (\partial^0 \bar C^{ij} + \partial^i \bar C^{j0} +  \partial^j \bar C^{0i})\, \beta_j 
=  2\, (\partial^0\, \bar{F}^i - \partial^i \, \bar F^0)\, \beta_i.
\end{eqnarray}
This term will be present in the nilpotent version of the BRST charge $Q_b^{(1)}$. 
Hence, this will be the {\it third} term in addition to the  {\it two} terms of (37). 
The BRST symmetry transformations on (40) leads to: $-\, 2\, (\partial^0\, \bar F^i - \partial^i\, \bar F^0)\, \partial_i\, C_2$
which should cancel out from some appropriate term of $Q_b$ when $s_b$ is applied on it. 
For this purpose, we modify the following: 
\begin{eqnarray}
-\, (\partial^0\, \bar\beta^i - \partial^i\, \bar\beta^0)\, \partial_i\, C_2 
= -\, 3 \, (\partial^0\, \bar\beta^i - \partial^i\, \bar\beta^0)\, \partial_i\, C_2
+ 2\, (\partial^0\, \bar\beta^i - \partial^i\, \bar\beta^0)\, \partial_i\, C_2. 
\end{eqnarray}
It is clear that when $s_b$ will act on the {\it second} term on the r.h.s. of (41), we shall get the
cancellation with: $-\, 2\, (\partial^0\, \bar F^i - \partial^i\, \bar F^0)\, \partial_i\, C_2$. 
Thus, the {\it second} term on the r.h.s. of (41) will be present as the {\it fourth} term in $Q_b^{(1)}$. 
We concentrate now on the {\it first} term of (41) which is present inside the integral and can be equivalently 
written as:  
\begin{eqnarray}
-\, 3\,  \int d^{D- 1}x\, (\partial^0\, \bar\beta^i - \partial^i\, \bar\beta^0)\, \partial_i\, C_2 
\equiv 3\,  \int d^{D- 1}x\, \partial_i\, (\partial^0\, \bar\beta^i - \partial^i\, \bar\beta^0)\,  C_2. 
\end{eqnarray}
We use the following EL-EoM\footnote{We would like point out that when a BRST-quantized theory is endowed
with a large number of fields, it becomes cumbersome to choose the EL-EoMs that ought to be used to obtain
the  off-shell nilpotent versions of the (anti-)BRST charges from their counterparts non-nilptent versions.
However, we have pointed out in our earlier work [22] that one has to start from the El-EoM for the gauge field in a given
BRST-quantized gauge theory. After that, it becomes obvious, in a sequence, which are the El-EoMs ought to
be used to obtain the off-shell nilpotent versions of the (anti-)BRST charges from the non-nilpotent
Noether (anti-)BRST charges. For instance, in our case,
we have used {\it only} the El-EoMs (32), (39) and (43).} from ${\cal L}_{B}$, namely; 
\begin{eqnarray}
\partial_\mu\, (\partial^\mu\, \bar\beta^\nu - \partial^\nu\, \bar\beta^\mu) - \partial^\nu\, B_2 = 0. 
\end{eqnarray}
With the choice: $\nu = 0$, we obtain the following equality: 
\begin{eqnarray}
-\, \partial_i\, (\partial^0 \, \bar\beta^i - \partial^i\, \bar\beta^0) =  \partial_0 \, B_2 \; \equiv \; \dot B_2.  
\end{eqnarray}
Thus, finally,  we obtain a BRST invariant quantity in the following explicit form: 
\begin{eqnarray}
3\,  \int d^{D- 1}x \; \,  \partial_i\, (\partial^0 \, \bar\beta^i - \partial^i\, \bar\beta^0)\, C_2
= -\, 3\,  \int d^{D- 1}x\, (\partial^0 \, B_2)\, C_2. 
\end{eqnarray}
This will be the {\it fifth} term of $Q_b^{(1)}$ in addition to {\it all}  the BRST invariant terms of our 
original Noether conserved BRST charge $Q_b$. Ultimately, we obtain the following expression for the conserved and off-shell nilpotent 
version of the BRST charge $[Q_b^{(1)}]$, namely;  
\begin{eqnarray}
Q_b^{(1)} &=&  \int d^{D- 1}x\, \Big[(\partial^0 \bar C^{ij} + \partial^i \bar C^{j0} 
+  \partial^j \bar C^{0i})\, (\partial_i\beta_j - \partial_j \beta_i) \nonumber\\
&-&  (\partial^0 B^{ij} + \partial^i B^{j0} + \partial^j  B^{0i})\, C_{ij}
+ 2\, (\partial^0 \bar F^i - \partial^i \bar F^0)\, \beta_i \nonumber\\
&+& 2\, (\partial^0 \, \bar\beta^i - \partial^i\, \bar\beta^0)\,\partial_i\,  C_2 
- \, 3\, C_2\, \partial^0\, B_2 + B_2\, \partial^0\, C_2 + B^{0i}\, f_i 
+ B_1\, f^0 \nonumber\\
&-&\, B\, \bar F^0  + (\partial^0 \, C^{ij} + \partial^i \, C^{j0} +  \partial^j \, C^{0i})\, B_{ij} 
 - (\partial^0\, \beta^i - \partial^i\, \beta^0)\, \bar F_i
\Big]. 
\end{eqnarray}
It is straightforward to check that  $s_b\, Q_b^{(1)} = -\, i\, \{Q_b^{(1)}, \, Q_b^{(1)} \} = 0$
(i.e. $[ Q_b^{(1)}]^2 = 0$) where the l.h.s. can be explicitly checked to be zero.

We dwell a bit, at this juncture, on the derivation of the off-shell nilpotent version of the 
anti-BRST charge [$Q_{ab}^{(1)}$] from the non-nilpotent Noether conserved anti-BRST charge
$Q_{ab}$ [cf. Eq. (26)]. Exactly as we have performed the systematic  exercise to obtain the off-shell nilpotent version of the 
BRST charge $[Q_b^{(1)}]$ from $Q_b$ [cf. Eq. (27)], we exploit the interplay of (i) the appropriate EL-EoMs 
from the Lagrangian density ${\cal L}_{\bar B}$, (ii) the Gauss divergence theorem, and (iii) the 
anti-BRST symmetry transformations (15) to obtain $Q_{ab}^{(1)}$ from $Q_{ab}$. We do {\it not} think it is
proper to repeat the similar kinds of computations as 
we have performed in the case of the BRST charge because it will {\it only}  be an academic exercise. Thus, we
straight away write  down the expression for the off-shell nilpotent version of the anti-BRST charge as:  
\begin{eqnarray}
Q_{ab}^{(1)} &=&  \int d^{D- 1}x\, \Big[(\partial^0 \bar B^{ij} + \partial^i \bar B^{j0} + \partial^j \bar B^{0i})\,\bar C_{ij} \nonumber\\
&-& (\partial^0 C^{ij} + \partial^i C^{j0} +  \partial^j  C^{0i})
(\partial_i \bar \beta_j -  \partial_j \bar \beta_i)
+ 2\, (\partial^0 \, F^i - \partial^i \,  F^0)\, \bar \beta_i  \nonumber\\
&+& 2\, (\partial^0 \, \beta^i - \partial^i\, \beta^0)\,\partial_i\, \bar C_2 
+ 3\, \bar C_2\, \partial^0\, B 
- B \, \partial^0\, \bar C_2  + \bar B^{0i}\, \bar f_i + B_2\, F^0 \nonumber\\
&+&   B_1\, \bar f^0  - (\partial^0 \,\bar  C^{ij} + \partial^i \,\bar  C^{j0} +  \partial^j \, \bar C^{0i})\, \bar B_{ij} 
- (\partial^0\,\bar \beta^i - \partial^i\, \bar \beta^0)\, F_i
\Big]. 
\end{eqnarray}
It is straightforward to check that $s_{ab}\, Q_{ab}^{(1)} = -\, i\, \{Q_{ab}^{(1)}, \, Q_{ab}^{(1)} \} = 0$ which 
implies that the anti-BRST charge $Q_{ab}^{(1)}$ is off-shell nilpotent [$(Q_{ab}^{(1)})^2 = 0$] of order two. 
The off-shell nilpotent versions of the (anti-)BRST charges are very essential and important as we shall see in  the 
next section. Besides it, as far as the BRST cohomology  is concerned, the nilpotency property
is very crucial and important (see, e.g.  [34] for details).

We wrap-up this section with the final remark that the non-nilpotent Noether conserved (anti-)BRST charges can be converted 
into the conserved and nilpotent (anti-)BRST charges by exploiting the interplay of (i) the appropriate EL-EoMs, 
(ii) the Gauss divergence theorem, and (iii) the applications of the (anti-)BRST symmetry transformations at appropriate 
places. It is interesting to point out that, primarily, it is the EL-EoMs that are used to obtain the off-shell nilpotent 
versions of the (anti-)BRST charges from the  non-nilpotent but conserved standard Noether 
(anti-)BRST charges. As a consequence, the resulting off-shell 
nilpotent versions of the (anti-)BRST charges are found to be conserved because the use of EL-EoMs and the 
Gauss divergence theorem would {\it not} violate the conservation law according to the basic principles 
behind the Noether theorem and ensuing conservation law (corresponding to the continuous symmetry transformations).   \\

\section{Physicality Criteria: Constraints at Quantum Level}

Unlike the {\it classical} starting Lagrangian density $({\cal L}_0)$ which is endowed with the first-class constraints:
$\Pi^{0ij} \approx 0, \; \partial_i\, \Pi^{ijk} \approx 0$, the (anti-)BRST invariant coupled 
(but equivalent) Lagrangian densities [cf. Eqs. (12), (13)] do {\it not} possess,  in their  explicit form, these first-class constraints. 
For instance, it can be readily checked that the following explicit expressions for the canonical conjugate momenta,
w.r.t. the gauge field $A_{\mu\nu\lambda}$,  namely; 
\begin{eqnarray}
\Pi^{\mu\nu\lambda} = \frac {\partial {\cal L}_B}{\partial (\partial_0 A_{\mu\nu\lambda})} 
= \frac {1}{3}\, H^{0\mu\nu\lambda} + \frac {1}{3}\, (\eta^{0\mu}\, B^{\nu\lambda} 
+ \eta^{0\nu}\, B^{\lambda\mu} + \eta^{0\lambda}\, B^{\mu\nu}), \nonumber\\
\Pi^{\mu\nu\lambda} = \frac {\partial {\cal L}_{\bar B}}{\partial (\partial_0 A_{\mu\nu\lambda})} 
= \frac {1}{3}\, H^{0\mu\nu\lambda} - \frac {1}{3}\, (\eta^{0\mu}\, \bar B^{\nu\lambda} 
+ \eta^{0\nu}\,\bar  B^{\lambda\mu} + \eta^{0\lambda}\, \bar B^{\mu\nu}), 
\end{eqnarray}
imply that the following relationships are true:  
\begin{eqnarray}
\Pi^{0ij} = \frac {1}{3}\,B_{ij}, \qquad \quad 
\Pi^{0ij} = -\, \frac {1}{3}\,\bar B_{ij}.
\end{eqnarray} 
Thus, we note that the original primary constraint (i.e. $\Pi^{0ij} \approx 0$) has been tracked with the space components
$(B_{ij}, \, \bar B_{ij})$ of the Nakanishi-Lautrup auxiliary fields $B_{\mu\nu}$ and $\bar B_{\mu\nu}$, respectively. 
In exactly similar fashion, we observe  that the following EL-EoMs from the Lagrangian densities 
${\cal L}_{ B}$ and ${\cal L}_{\bar B}$ [cf. Eqs. (12), (13)], respectively, are as follows: 
\begin{eqnarray}
\partial_\mu H^{\mu\nu\lambda\xi} + (\partial^\nu B^{\lambda\xi} 
+ \partial^\lambda B^{\xi\nu} +  \partial^\xi B^{\nu\lambda}) = 0, \nonumber\\
\partial_\mu H^{\mu\nu\lambda\xi} - (\partial^\nu \bar B^{\lambda\xi} 
+ \partial^\lambda \bar B^{\xi\nu} +  \partial^\xi \bar B^{\nu\lambda}) = 0.
\end{eqnarray}
Making the choices: $\nu = 0, \lambda = j$ and $\xi = k$, we obtain the following 
\begin{eqnarray}
\partial_i H^{i0jk} + (\partial^0 B^{jk} + \partial^j B^{k0} +  \partial^k B^{0j}) = 0, \nonumber\\
\partial_i H^{i0jk} - (\partial^0 B^{jk} + \partial^j B^{k0} +  \partial^k B^{0j}) = 0,
\end{eqnarray}
where we have taken $H^{00jk} = 0$ (i.e. strongly equal to zero) at the 
{\it quantum} level of our discussion because the coupled (but equivalent) Lagrangian densities 
${\cal L}_{ B}$ and ${\cal L}_{\bar B}$,  do {\it not} possess any component of 
momenta (w.r.t. the gauge field $A_{\mu\nu\lambda}$) equal to zero. 
Taking into account the expression for the secondary constraint ($\partial_i\, \Pi^{ijk} \approx 0$) 
of the original {\it classical} gauge theory, we find that (51) leads to the 
following explicit relationships: 
\begin{eqnarray}
&&(\partial^0 B^{jk} + \partial^j B^{k0} +  \partial^k B^{0j}) = 3\, \partial_i\, \Pi^{ijk}, \nonumber\\
&&(\partial^0 B^{jk} + \partial^j B^{k0} +  \partial^k B^{0j}) = -\, 3\, \partial_i\, \Pi^{ijk}.
\end{eqnarray}
The above equations [cf. Eq. (52)] demonstrate that the secondary constraint ($\partial_i\, \Pi^{ijk} \approx 0$) of the original 
{\it classical} gauge theory has been traded with the specific combinations of the derivatives on the 
Nakanishi-Lautrup type auxiliary fields that are present on the l.h.s. of equation 
(52). According to Dirac's quantization method of theories that are endowed with constraints, it is an essential requirement that the operator form of these constraints must annihilate the physical states (i.e. $| phys >$)
at the quantum level. For our massless D-dimensional free  Abelian 3-form theory, this 
essential requirement, according to Dirac's quantization method,  implies that the following conditions 
\begin{eqnarray}
\Pi^{0ij} \, |\,  phys > = 0, \qquad \partial_i\, \Pi^{ijk} \, |\, phys > = 0,
\end{eqnarray}
must be satisfied for the well-defined  quantum theory.
At this stage, it will be noted that the physical states (i.e. $ |\, phys > $) of our Abelian 3-form gauge theory correspond to only the gauge fields 
(and their canonical conjugate momenta).

The stage is now set to capture the Dirac-quantization conditions (53) within the framework of BRST formalism 
where the off-shell nilpotent versions of the (anti-)BRST charges (i.e. $Q_{(a)b}^{(1)}$) play a very important
role. The nilpotency property is very crucial when we discuss the physical states (i.e. $| phys >$) because the 
gauge transformed states can be shown to be the {\it exact} states w.r.t. the nilpotent 
versions of the (anti-)BRST charges. The nilpotency property 
ensures that such states are {\it trivial} as far as the (anti-)BRST charges are concerned (see, e.g. [34]).
The physicality criteria, w.r.t. the nilpotent (anti-)BRST charges [$Q_{(a)b}^{(1)}$], are as follows
(see, e.g. [14,15,31,34,35] for details): 
\begin{eqnarray}
Q_{b}^{(1)} \, |\, phys > = 0, \qquad \qquad Q_{ab}^{(1)} \, |\, phys > = 0.
\end{eqnarray}
In our BRST-quantized free massless Abelian 3-form theory, the physical gauge fields (as well as the associated Nakanishi-Lautrup auxiliary fields)
and the (unphysical) (anti-)ghost fields (as well as the associated auxiliary (anti-)ghost fields)
are decoupled. In other words, there are {\it no} interaction terms between {\it them} in our entire BRST-quantized
free massless Abelian 3-form theory
that is described by the coupled (anti-)BRST invariant Lagrangian densities [cf. Eqs. (12),(13)].
Right from the beginning, the (anti-)ghost fields are redundant in some sense. Thus, the total inner product
space of the states of our BRST-quantized theory is precisely the tensor product of the inner product spaces of the states that are generated 
by the action of  all the physical fields as well as the (anti-)ghost fields on their respective vacua. In other words,
when the polynomials of the physical as well as the (anti-)ghost fields operate on their respective vacua, we obtain, ultimately, the
inner product of the spaces of the physical states i.e. $| phys >$) and the ghost states. In physical terms, it boils down to the
statement that the 
total quantum Hilbert space of states is the inner product of the physical states and the ghost states where the physical states 
(i.e. $ |\, phys > $), as is obvious, carry the ghost number equal to zero. 
It is self-evident that the (anti-)ghost field 
operators of the conserved and nilpotent versions of the (anti-)BRST charges [cf. Eqs. (46),(47)] will act on the
ghost states in the total quantum Hilbert space of states which will be, obviously, 
{\it not} equal to zero.
Thus, the physicality criteria [cf. Eq. (54)],  w.r.t. the 
conserved and nilpotent versions of the (anti-)BRST charges $ Q_{(a)b}^{(1)}$, will require that the 
field operators with the ghost number {\it zero} (which are the coefficients of the independent 
{\it basic} (anti-)ghost fields and/or the derivative(s) on them)
would operate on the physical states to yield {\it zero}. In other words, first of all, if we concentrate on the
the above physicality criterion w.r.t. the nilpotent 
BRST charge $Q_{b}^{(1)} $ [cf. Eq. (46)], 
we obtain the following quantization conditions 
(in terms of the first-class constraints) on  the physical states (i.e. $ |\, phys > $)  
of the {\it total} quantum Hilbert space of states of our theory, namely; 
\begin{eqnarray}
B_{ij}\,  |\, phys > = 0  &\Longrightarrow&    \Pi^{0ij} \, |\,  phys > = 0, \nonumber\\
(\partial^0 B^{jk} + \partial^j B^{k0} +  \partial^k B^{0j}) |\, phys > = 0 &\Longrightarrow& 
\partial_i\, \Pi^{ijk} \, |\, phys > = 0. 
\end{eqnarray}
A close look at the nilpotent version of the BRST charge [cf. Eq. (46)] shows that the Nakanishi-Lautrup auxiliary field $B_{ij} $
is the coefficient of the totally antisymmetric combination (i.e. $\partial^0 \, C^{ij} + \partial^i \, C^{j0} +  \partial^j \, C^{0i}$)
of the derivatives on the independent {\it basic} ghost fields and the totally antisymmetric combination (i.e. $\partial^0 B^{ij} 
+ \partial^i B^{j0} + \partial^j  B^{0i}$) of the derivatives on the Nakanishi-Lautrup auxilairy fields
is the coefficient of the independent {\it basic} ghost field $C_{ij}$. Hence, the physicality criterion
w.r.t. the nilpotent BRST charge $Q_b^{(1)}$ leads to the observations in the above equation (55) because the auxiliary field $B_{ij} $
and the specific combination of the cyclic derivatives on the Nakanishi-Lautrup auxiliary fields (i.e. $\partial^0 B^{ij} 
+ \partial^i B^{j0} + \partial^j  B^{0i}$), carrying the ghost number equal to zero,  are connected with the constraints 
that are associated with the gauge field (which is the {\it physical} field in our theory). It is interesting to point out that the
terms like: $B_2 \,\partial^0\, C_2, - 3\, C_2\, \partial^0\, B_2$, etc., which are {\it also} associated with the independent basic ghost field $C_2$
in equation (46), do {\it not} contribute to the quantization conditions on the physical states (in the physicality criterion
w.r.t. $Q_b^{(1)}$)  because the auxiliary field $B_2$ carries a ghost number equal to (-2).

On the other hand, the physicality criterion w.r.t. the off-shell nilpotent version of the 
anti-BRST charge $Q_{ab}^{(1)} $ [cf. Eq. (47)] implies that we have the following quantization 
conditions on the physical states of our theory:
\begin{eqnarray}
\bar B_{ij}\,  |\, phys > = 0  &\Longrightarrow&   \Pi^{0ij} \, |\,  phys > = 0, \nonumber\\
(\partial^0 \bar B^{jk} + \partial^j \bar B^{k0} +  \partial^k \bar B^{0j}) |\, phys > = 0  &\Longrightarrow&  
\partial_i\, \Pi^{ijk} \, |\, phys > = 0. 
\end{eqnarray}
In other words, the coefficients of the combination (i.e. $\partial^0 \, \bar C^{ij} + \partial^i \, \bar C^{j0} +  \partial^j \, \bar C^{0i}$)
and the basic anti-ghost field $\bar C_{ij}$, carrying the ghost numbers equal to zero,
have annihilated the physical states (existing in the total Hilbert space of states).
It should be noted that the physical states (i.e. $ |\, phys > $) do {\it not} acquire any {\it non-zero} ghost number after the 
operation of the Nakanishi-Lautrup type auxiliary fields $\bar B_{ij}$ and the specific combination of the derivatives on these
auxiliary fields (i.e. 
$\partial^0 \bar B^{jk} + \partial^j \bar B^{k0} +  \partial^k \bar B^{0j}$), respectively. 
It is an interesting observation that the terms like: $-\, B\,\partial^0\, {\bar C}_2, 3\, \bar C_2\,\partial^0\, B$, etc., do {\it not} contribute
to the quantization conditions on the physical states [in the physicality criterion w.r.t. (47)] because the
auxiliary field $B$ carries the ghost number (+2).
Thus, ultimately, we note that the Dirac-quantization conditions are satisfied with the nilpotent 
(anti-)BRST charges in the case of our theory where the constraints are: 
$\Pi^{0ij} \approx 0, \;  \partial_i\, \Pi^{ijk} \approx 0$ at the classical level.  
We lay emphasis on the fact that the standard Noether conserved charges $Q_{(a)b}$
[cf. Eqs. (26), (27)] can {\it not} 
lead to the annihilation of the physical states by the operator forms of {\it both} the first-class constraints. 
To be precise, we obtain the quantization condition: $B_{ij}\, |phys> = 0$ and/or  $\bar B_{ij}\, |phys> = 0$
which corresponds to the annihilation of the physical states {\it only} by the operator form of the 
primary constraint of our {\it classical} D-dimensional massless Abelian 3-form gauge theory.

We lay stress on the observations (55) and (56) which imply
that the nilpotent BRST and anti-BRST charges lead to the {\it same} 
quantization conditions on the physical states as far as the first-class constraints of the original {\it classical} gauge theory are concerned.
We would like to mention here, in passing, that all the rest of the terms of the nilpotent versions of the 
(anti-)BRST charges  $Q_{(a)b}^{(1)}$ are either connected with the basic (anti-)ghost fields 
[e.g. $2\, (\partial^0\, \bar F^i - \partial^i\, \bar F^0)\, \beta_i$ in (46)]
and/or with the auxiliary (anti-)ghost fields 
[e.g. $-\, B\, F^0$ in (46)]
that carry individually some non-zero ghost numbers. Hence, these terms would {\it not} lead
to any quantization conditions on the physical states. The exceptions are the pairs of
terms ($B^{0i}\, f_i,\; B_1 \,f^0$) and ($\bar B^{0i}\, \bar f_i, \; B_1\, \bar f^0$) that are present in the
expressions for the nilpotent BRST charge [cf. Eq. (46)] and the anti-BRST charge [cf. Eq. (47)], respectively.
A noteworthy point, at this stage, is one of the key observations that these pairs ($\bar B^{0i}\, \bar f_i, \; B_1\, \bar f^0$) and 
($B^{0i}\, f_i,\; B_1 \,f^0$) are basically associated with the auxiliary (anti-)ghost fields $(\bar f_\mu)f_\mu$
which are {\it not} independent (in the true sense of the word) as they participate in the specific kinds of the (anti-)BRST invariant
CF-type restrictions on our theory. In the next paragraph,
we comment on these specific terms and explain the reasons behind their {\it nil} contributions to the quantization conditions
on the physical states.

We end this section with the following concluding remark. We point out that we have {\it not} taken 
($B^{0i}\, |phys> = 0,\, B_1 |phys> = 0$) and/or ($\bar B^{0i}\, |phys> = 0,\, B_1 |phys> = 0$) on the physical states because (i) the  
pairs of the auxiliary fields ($B^{0i}, B_1$) and/or ($\bar B^{0i}, B_1$)
are {\it not} associated with the independent {\it basic} 
(anti-)ghost fields of our theory (even though they carry the ghost numbers equal to zero), 
and (ii) we further observe that the canonical conjugate momenta [i.e. $\Pi^{\mu}_{(\phi)}$] w.r.t. the Lorentz-vector
$\phi_\mu$ field, from the Lagrangian densities  ${\cal L}_{ B}$ and ${\cal L}_{\bar B}$, are connected with the pairs
of auxiliary fields ($B^{0i}, B_1$) and/or ($\bar B^{0i}, B_1$) because we have the following:
\begin{eqnarray}
\Pi^{\mu}_{(\phi)} = \frac {\partial {\cal L}_B}{\partial (\partial_0 \phi_{\mu})}
= B^{0\mu} + \eta^{0\mu}\, B_1,  \nonumber\\
\Pi^{\mu}_{(\phi)} = \frac {\partial {\cal L}_{\bar B}}{\partial (\partial_0 \phi_{\mu})}
= \bar B^{0\mu} + \eta^{0\mu}\,  B_1. 
\end{eqnarray}
The above equation shows that the pairs ($B^{0i}, B_1$) and/or ($\bar B^{0i}, B_1$) are {\it not} a set of constraints on our massless 
Abelian 3-form theory (at the classical level). Rather, they are equal to $\Pi^i_{(\phi)} = $ $B^{0i}$ or $\bar B^{0i}$ and $\Pi^0_{(\phi)} = B_1$. 
However, the field $\phi_\mu$ has appeared {\it only} at the quantum level and there is {\it no} trace of it at the
{\it classical} level. To be precise, there are {\it no} constraints at the quantum level 
for the coupled (anti-)BRST invariant Lagrangian densities (when our {\it classical} gauge theory is properly BRST-quantized).
Hence, the pairs ($B^{0i}, B_1$) and/or ($\bar B^{0i}, B_1$), even though endowed with the ghost numbers equal to zero, 
can {\it not} impose any condition(s) on the physical states (i.e. $|phys> $)
of our theory because they are {\it not}  associated with the 
independent basic (anti-)ghost fields in the expressions for the 
nilpotent versions of the (anti-)BRST charges.  Furthermore, we would like to point out that the pair of terms 
($\bar B^{0i}\, \bar f_i, \; B_1\, \bar f^0$) and ($B^{0i}\, f_i,\; B_1 \,f^0$),  in the expressions for the nilpotent versions
of the (anti-)BRST charges,  are associated with the specific components of the auxiliary (anti-)ghost fields $(\bar f_\mu)f_\mu$ which are {\it not} independent 
as they are constrained to obey the 
(anti-)BRST invariant CF-type restrictions: $\bar f_\mu + \bar F_\mu = \partial_\mu \bar C_1, \, f_\mu + F_\mu = \partial_\mu C_1 $.
This is the key reason behind our claim that the physicality criteria w.r.t. $Q_{(a)b}^{(1)}$ would 
{\it not} lead to the quantization conditions: $\bar B^{0i} |phys> = 0, \,  B^{0i} |phys> = 0$ and $B_1 |phys> = 0$. \\

\section{Conclusions}

In our present endeavor, the emphasis is laid on the  constraint structures of the D-dimensional free massless Abelian 3-form theory at 
the {\it classical} as well as at the {\it quantum} level. For the {\it latter} case, we have exploited the 
theoretical beauty and strength of the BRST formalism. At the {\it classical} level, the starting  Lagrangian density
[cf. Eq. (1)] is endowed with the primary and secondary constraints that are found to belong to the first-class 
variety within the ambit of Dirac's classification scheme of constraints [11-13]. 
For the BRST-quantized theory, these constraints appear in their operator form when we demand that the physical states 
of the quantum theory are {\it those} that are annihilated by the conserved and off-shell nilpotent versions of the (anti-)BRST
charges. This observation is consistent with the Dirac-quantization scheme for theories that are 
endowed with any kind of constraints [11-15] because the operator form of these constraints must annihilate the 
physical state. This statement is mathematically captured in 
the requirement $Q_{(a)b} \, | phys > = 0$ within the framework of BRST formalism where the conserved and nilpotent (anti-)BRST
charges are denoted by  $Q_{(a)b}$ for this purpose.

One of the interesting observations of our present endeavor  is the fact that the standard Noether theorem does not lead to the 
derivations of the off-shell nilpotent (anti-)BRST charges even though they are derived from the 
continuous, infinitesimal and  off-shell nilpotent (anti-)BRST symmetry transformations. In our earlier work [22],
we have developed a systematic theoretical method to obtain the off-shell nilpotent versions of the 
(anti-)BRST charges from the non-nilpotent Noether (anti-)BRST charges. In a recent work [27], we have shown that
our method is also applicable in the context of the (anti-)co-BRST charges. In our present investigation, 
we have exploited our earlier proposal [22] to obtain the off-shell nilpotent versions of the (anti-)BRST charges which are very
useful in the context of the physicality criteria (cf. Sec. 6) where we have demonstrated the appearance of the 
operator form of the first-class constraints of the original {\it classical} theory. In fact, the operator form of the 
first-class constraints are found to annihilate the physical states at the {\it quantum} level which is consistent with the Dirac 
quantization scheme applied to theories (endowed with any kind of constraints). Thus, we have been able to 
discuss the first-class constraints of our theory at the {\it classical} level as well as at the {\it quantum} level (within the framework
of the BRST formalism through the physicality criteria w.r.t. the off-shell nilpotent versions of the conserved (anti-)BRST charges).

One of the key signatures of a BRST-quantized theory is the existence of the CF-type restriction(s). In our present case,
there are {\it three} such restrictions [cf. Eq. (20)]. In our present endeavor, we have laid a bit of emphasis on the derivation 
of the CF-type restrictions from the EL-EoMs [cf. Eq. (32)] that have been derived from the coupled (but equivalent) 
(anti-)BRST invariant Lagrangian densities  ${\cal L}_{ B}$ and ${\cal L}_{\bar B}$, respectively. We have {\it also}
pointed out  the existence of the CF-type restrictions in the context of (i) the absolute anticommutativity 
of the (anti-)BRST symmetry transformations [cf. Eqs. (19) (20)], and (ii) the proof of the 
equivalence [28] of the coupled Lagrangian densities  ${\cal L}_{ B}$ and ${\cal L}_{\bar B}$ from the 
point of view of (anti-)BRST symmetry transformations (cf. footnote on Page 7). For the BRST-quantized theories, 
the esistence of the CF-type restrictions is as fundamental as the existence of the first-class constraints 
in the context of {\it classical} gauge theories. The CF-type restrictions have been shown to be connected with the 
geometrical objects called gerbes [28, 36] which,  primarily, prove the independent identity of the BRST and anti-BRST symmetry  
transformations and their corresponding charges because of their absolute anticommutativity properties.  
The absolute anticommutativity  property of the nilpotent (anti-)BRST charges has been proven in our Appendix A and the existence of the 
CF-type restrictions has been demonstrated.

We would like to make a few brief comments on the physicality criteria (cf. Sec. 6) w.r.t. the conserved and nilpotent versions of the
(anti-)BRST charges where it has been shown explicitly that the quantization conditions [cf. Eqs. (55),(56)]
on the physical states (i.e. $| phys >$)
of the BRST-quantized theory are  obtained from (i) the field operators (of the above nilpotent (anti-)BRST charges) 
that are endowed with the ghost numbers equal to zero, and (ii)
these ghost number zero operators are {\it either} the coefficients of the independent {\it basic} (anti-)ghost fields {\it or} the derivatives on the 
{\it latter} fields which do not participate in any kind of the (anti-)BRST invariant CF-type restrictions. It has also been demonstrated that
a few {\it zero} ghost number operators do {\it not} contribute 
[cf. Eq. (57) and ensuing discussions] to the quantization conditions on the physical states (i.e. $| phys >$) if they happen 
to be the coefficients of the auxiliary (anti-)ghost fields (or the derivatives on them). To be more precise, the auxiliary (anti-)ghost fields,
whose coefficients are the field operators with the ghost number equal to zero, would not contribute to the quantization conditions on the
physical states, even if, the auxiliary (anti-)ghost fields participate or do {\it not} participate in the (anti-)BRST
invariant CF-type restrictions of our theory. This statement is sacrosanct in the context of the physicality criteria w.r.t. the conserved
and nilpotent versions of the (anti-)BRST charges of a properly D-dimensional BRST-quantized Abelian $p$-form 
massless and/or St$\ddot u$ckelberg-modified massive gauge theory [26].

We plan to pursue our study of the higher $p$-form gauge theories up to, at least,  
$p = 5$ form theories as they are relevant to the self-dual superstring theories that are consistently defined in the
$D = 10$ dimensions of spaetime. In other words, we know that up to $p = 5$ form gauge fields can exist in the $D = 10$
dimensions of spacetime. The BV formalism of the Abelian 3-form theory has been done in superspace [37] where the 
extended (anti-) BRST symmetries (including the shift symmetry) have been obtained for specific Lagrangian densities.
It will be nice to apply our present insights and understandings in the context of these Lagrangian densities, too.
Furthermore, the knowledge gained in our present endeavor will be useful in the 
completion of earlier work on such analysis and discussion for the St$\ddot u$ckelberg-modified 
massive Abelian 3-form theory [26] which is yet to be completed. It is gratifying to state that
this work has been completed now with the help of our experience and understanding of our present endeavor.
The Abelian 3-form theory has natural connection with the M-theory where the BRST analysis of the ABJM model has been carried out
(see, e.g. [38]). We can also study ABJM theory in the light of our present endeavor
where the emphasis has been laid on the constraint structure.\\

\vskip 0.5 cm

\noindent
{\bf Acknowledgments}\\

\noindent
One of us (AKR) thankfully acknowledges the financial support from the 
Institution of Eminence (IoE) {\it Research Incentive Grant} of PFMS (Scheme No. 3254-World Class Institutions)
to which BHU, Varanasi, belongs. The present investigation has been carried out under the above
{\it Research Incentive Grant} which has been launched by the Govt. of India for the promotion of research activities in our country. 
Thanks are also due to our esteemed {\bf Reviewer} for fruitful and enlightening comments on our Sec. 6 due to which our presentation
of this section has certainly become more transparent and readable. \\

\vskip 0.6cm

\noindent
{\bf Conflicts of Interest}\\

\vskip 0.6cm

\noindent
The authors declare that there are no conflicts of interest. \\

\vskip 0.6cm

\noindent
{\bf Data Availability}\\

\noindent
No data were used to support this study.\\

\vskip 0.7 cm 
\begin{center}
{\bf Appendix A: \bf On the CF-Type Restrictions}\\
\end{center}
\noindent
The existence of the (non-)trivial CF-type restrion(s) is one of the decisive features of the BRST-quantized theory. 
We have been able to demonstrate the existence and importance of the CF-type restrictions from 
(i) the absolute anticommutativity of the (anti-)BRST symmetries [cf. Eqs (15), (16), (19)], 
and (ii) the  EL-EoMs from the coupled (but equivalent) Lagrangian densities 
${\cal L}_{ B}$ and ${\cal L}_{\bar B}$ [cf. Eqs. (22), (23), (24)]. The purpose of our 
present Appendix is to demonstrate the existence of the {\it three} CF-type restrictions 
(i.e $B_{\mu\nu} + \bar B_{\mu\nu} = (\partial_\mu\, \phi_\nu - \partial_\nu\, \phi_\mu), \; 
f_\mu + F_\mu = \partial_\mu\, C_1$ and $\bar f_\mu + \bar F_\mu = \partial_\mu\, \bar C_1$)
from the requirement of the absolute anticommutativity of the off-shell nilpotent versions 
of the (anti-)BRST charges [$Q_{(a)b}^{(1)}$] where the idea behind the continuous symmetry transformations 
and their generators as the Noether conserved charges has been exploited. Mathematically, we wish to prove that
the following is true  
\[
s_{ab}\, Q_b^{(1)} = -\, i\, \{Q_b^{(1)}, \, Q_{ab}^{(1)} \} = 0 \Longrightarrow
Q_b^{(1)}\, Q_{ab}^{(1)} + Q_{ab}^{(1)}\, Q_{b}^{(1)} = 0, 
\eqno (A.1)
\]
where the l.h.s. can be computed by exploiting the anti-BRST symmetry transformations (15) and the explicit 
expression for the off-shell nilpotent BRST charge $Q_{b}^{(1)}$ [cf. Eq. (46)]. It turns out that the 
direct application of (16) on (46) leads to the following: 
\[
s_{ab}\, Q_b^{(1)} = \int d^{D-1}\, x \Big[  
(\partial^0\, \bar B^{ij} + \partial^i\, \bar B^{j0} + \partial^j\, \bar B^{0i})\, B_{ij}
- (\partial^0\, B^{ij} + \partial^i\, B^{j0}
\]
\[
+ \, \partial^j\, B^{0i})\, \bar B_{ij} - (\partial^0\, \bar C^{ij} 
+ \partial^i\, \bar C^{j0} + \partial^j\, \bar C^{0i})\,(\partial_i\, F_j - \partial_j\, F_i) 
\]
\[
~~~~~~~~~~~- (\partial^0\,  C^{ij} + \partial^i\, C^{j0} + \partial^j\, C^{0i})\,(\partial_i\, \bar f_j -
 \partial_j\, \bar f_i) + 2\, (\partial^0\, \bar\beta^i - \partial^i\, \bar\beta^0)\,\partial_i\,B
 \]
 \[
 ~~~~~~~~~- 2\, (\partial^0\, \bar F^i - 
\partial^i\, \bar F^0)\, F_i +  B_2\, \partial^0\, B - 2\, B\, \partial^0\, B_2 
+ (\partial^0\, \bar f^i - \partial^i\, \bar f^0)\, f_i
\]
\[
 ~~~~~~~~~~~~~~~~~~~~- B^{0i}\, \partial_i\, B_1 - 
B_1\, \partial^0 \, B_1 + (\partial^0\, \beta^i    - \partial^i\, \beta^0)\, \partial_i\, B_2 
- (\partial^0\,  F^i - \partial^i\, F^0)\, \bar F_i 
\Big].
\eqno (A.2)
\]
It is pertinent to point out, at this stage,  that we can exploit the EL-EoMs as well as Gauss's divergence
theorem on the r.h.s. of the above equation. In other words, we have the following: 
\[
\int d^{D-1}\, x \Big[2\, (\partial^0\, \bar\beta^i - \partial^i\, \bar\beta^0)\,\partial_i\,B - B^{0i}\, \partial_i\, B_1
+ (\partial^0\, \beta^i - \partial^i\, \beta^0)\, \partial_i\, B_2 \Big]
\]
\[
\equiv \int d^{D-1}\, x \Big[-\, 2\,\partial_i\, (\partial^0\, \bar\beta^i - \partial^i\, \bar\beta^0)\, B 
+   (\partial_i\, B^{0i})\, B_1 - \partial_i\,(\partial^0\, \beta^i - \partial^i\, \beta^0)\, B_2   \Big], 
\eqno (A.3)
\]
where we have dropped the total space derivative terms. Using the following EL-EoMs, namely; 
\[
\partial_\mu\, B^{\mu\nu} + \partial^\nu\, B_1 = 0, \quad 
\partial_\mu\, (\partial^\mu\, \bar \beta^\nu - \partial^\nu\, \bar\beta^\mu) 
- \partial^\nu\, B_2 = 0, 
\]
\[
 \partial_\mu\, (\partial^\mu\,  \beta^\nu - \partial^\nu\,\beta^\mu) 
+ \partial^\nu\, B = 0, 
\eqno (A.4)
\]
it is evident that we have the following result: 
\[
\int d^{D-1}\, x \Big[
(\partial^0\, \beta^i - \partial^i\, \beta^0)\, \partial_i\, B_2  - B_1\, \partial^0\, B_1 
\]
\[
~~~~~~~~~~+ 2\, (\partial^0\, \bar\beta^i - \partial^i\, \bar\beta^0)\,\partial_i\,B 
 + B_2\, \partial^0\, B -2\, B\, \partial^0\, B_2 
\Big] = 0. 
\eqno (A.5)
\]
In a similar fashion, using the Gauss divergence theorem and antisymmetric property, we have the following equality: 
\[
-\, \int d^{D-1}\, x \Big[
(\partial^0\, \bar C^{ij} + \partial^i\, \bar C^{j0} + \partial^j\, \bar C^{0i})\,(\partial_i\, \beta_j - \partial_j\, \beta_i) 
\]
\[
~~~~~~~~~~- (\partial^0\,  C^{ij} + \partial^i\, C^{j0} + \partial^j\, C^{0i})\,(\partial_i\, \bar f_j -\partial_j\, \bar f_i) \Big]
\]
\[
\equiv 2\, \int d^{D-1}\, x \Big[
\partial_i\,(\partial^0\, \bar C^{ij} + \partial^i\, \bar C^{j0} + \partial^j\, \bar C^{0i})\ \beta_j   ~~~~~~~~~~
\]
\[
+ \partial_i\, (\partial^0\,  C^{ij} + \partial^i\, C^{j0} + \partial^j\, C^{0i})\,\bar f_j
\Big].
\eqno (A.6)
\]
We exploit the theoretical usefulness of the following EL-EoMs
\[
\partial_\mu\, (\partial^\mu\, \bar C^{\nu\lambda} + \partial^\nu\, \bar C^{\lambda\mu} + \partial^\lambda\, \bar C^{\mu\nu})
- \frac{1}{2}\, (\partial^\nu\, \bar{F}^\lambda - \partial^\lambda \, \bar F^\nu) = 0, 
\]
\[
\partial_\mu\, (\partial^\mu\,  C^{\nu\lambda} + \partial^\nu\,  C^{\lambda\mu} + \partial^\lambda\,  C^{\mu\nu})
+ \frac{1}{2}\, (\partial^\nu\, {f}^\lambda - \partial^\lambda \,  f^\nu) = 0, 
\eqno (A.7)
\]
to recast the r.h.s. of (A.6) as: 
\[
\int d^{D-1}\, x \Big[(\partial^0\, \bar F^i - \partial^i\, \bar F^0)\, F_i
+ (\partial^0\,  f^i - \partial^i\,  f^0)\, \bar f_i
\Big].
\eqno (A.8)
\]
Ultimately, at this juncture, we have the following explicit expression (taking into account (A.5) and (A.8))  for the 
quantity $s_{ab}\, Q_b^{(1)}$: 
\[
s_{ab}\, Q_b^{(1)} = \int d^{D-1}\, x \Big[  
(\partial^0\, \bar B^{ij} + \partial^i\, \bar B^{j0} + \partial^j\, \bar B^{0i})\, B_{ij}
- (\partial^0\, B^{ij} + \partial^i\, B^{j0} 
\]
\[
+ \partial^j\, B^{0i})\, \bar B_{ij} 
+ (\partial^0\,  f^i - \partial^i\,  f^0)\,\bar  f_i
+  (\partial^0\,  \bar f^i - \partial^i\,  \bar f^0)\, f_i  ~~
\]
\[
- (\partial^0\,  F^i - \partial^i\, F^0)\, \bar F_i
- (\partial^0\, \bar F^i - \partial^i\, \bar F^0)\, F_i
\Big].   ~~~~~~~~~~~~~
\eqno (A.9)
\]
We are now at the position to show the existence of the CF-type restrictions on the r.h.s. of the above equation.

First of all, we note that the first {\it two} terms of (A.9) can be expressed as
\[
\partial^0 \big\{\bar B^{ij}  + B^{ij} - (\partial^i\, \phi^j - \partial^j\, \phi^i)\big\}\, B_{ij} ~~~~~~~~~~~~~~~~~~~~~~~~~~~~~~~~~~~
\]
\[
+\, \partial^i\, \big\{\bar B^{j0}  + B^{j0} - (\partial^j\, \phi^0 - \partial^0\, \phi^j)\big\}\, B_{ij}  ~~~~~~~~~~~~~~~~~~~~~~~~~~~~~~~~~ 
\]
\[
+ \,\partial^j\, \big\{\bar B^{0i}  + B^{0i} - (\partial^0\, \phi^i - \partial^i\, \phi^0)\big\}\, B_{ij}  ~~~~~~~~~~~~~~~~~~~~~~~~~~~~~~~~~~
\]
\[
- \; (\partial^0\, B^{ij} + \partial^i\, B^{j0} + \partial^j\, B^{0i})
\big[B_{ij} 
+ \bar B_{ij} - (\partial_i\, \phi_j - \partial_j\, \phi_i)\big]  ~~~~~~~~~
\]
\[
- (\partial^0\, B^{ij} + \partial^i\, B^{j0} + \partial^j\, B^{0i})\,
(\partial_i\, \phi_j - \partial_j\, \phi_i), ~~~~~~~~~~~~~~~~~~~~~~~~~~
\eqno (A.10)
\]
where we note the presence of the CF-type restriction: 
$B_{\mu\nu} + \bar B_{\mu\nu} = (\partial_\mu\, \phi_\nu - \partial_\nu\, \phi_\mu)$
in all the terms except the {\it last} one which can be re-expressed as 
\[
- 2\, (\partial^0\, B^{ij} + \partial^i\, B^{j0} + \partial^j\, B^{0i})
\partial_i\, \phi_j 
\; \; \equiv \; \; + 2\, \partial_i\, (\partial^0\, B^{ij} + \partial^i\, B^{j0} + \partial^j\, B^{0i})\, \phi_j, 
\eqno (A.11)
\]
where we have taken care of the integration and dropped a total space derivative term due to 
Gauss's divergence theorem. Using the EL-EoM: 
\[
\partial_\mu\, H^{\mu\nu\lambda\zeta} + (\partial^\nu\, B^{\lambda\zeta} + \partial^\lambda\, B^{\zeta\nu}
+ \partial^\zeta\, B^{\nu\lambda}) = 0, 
\eqno (A.12)
\]
we note that the r.h.s. of (A.11) turns out to be
\[
2\, \partial_i\, (\partial_k\, H^{0kij} )\, \phi_j = 0,
\eqno (A.13)
\]
because the sum of the antisymmetric and symmetric indices is always zero. Now we concentrate on the left-over terms
of (A.9) where we can re-express: 
\[
(\partial^0\,  f^i - \partial^i\,  f^0)\, \bar f_i = [\partial^0\, (f^i + F^i - \partial^i\, C_1)
\]
\[
- \partial^i\, (f^0 + F^0 - \partial^0\, C_1)]\bar f_i 
- (\partial^0\,  F^i - \partial^i\,  F^0)\, \bar f_i. 
\eqno (A.14)
\]
We note that in the first two terms on the r.h.s.,  we have the presence of the 
CF-type restriction: $f_\mu + F_\mu = \partial_\mu\, C_1$. Thus, we have the following
\[
(\partial^0\,  f^i - \partial^i\,  f^0)\, \bar f_i 
 - (\partial^0\,  F^i - \partial^i\,  F^0)\, \bar F_i 
= \partial^0\, (f^i + F^i - \partial^i\, C_1)
- \partial^i\, (f^0 + F^0 - \partial^0\, C_1) 
\]
\[
- (\partial^0\,  F^i - \partial^i\,  F^0)\, (\bar f_i  + \bar F_i - \partial_i\,\bar C_1) 
  (\partial^0\,  F^i - \partial^i\,  F^0)\,\partial_i\, \bar C_1,
\eqno (A.15)
\]
where we have the presence of the CF-type  restriction: $f_\mu + F_\mu = \partial_\mu\, C_1$, in all the 
terms except the last term. Taking into account the Gauss divergence theorem and integration, 
we have the last term as 
\[
(\partial^0\,  F^i - \partial^i\,  F^0)\, \partial_i\, \bar C_1 
= \partial_i\, (\partial^0\,  F^i - \partial^i\,  F^0)\, \bar C_1. 
\eqno (A.16)
\]
Using one of the appropriate EL-EoMs from  (22), it can be checked that the r.h.s. of the above equation can be written as
\[
2\, \partial_i\, \partial_j [\partial^0\, C^{ij} + \partial^i\, C^{j0} + \partial^j\, C^{0i}]\, \bar C_1 = 0, 
\eqno (A.17)
\]
because there is sum of the antisymmetric and symmetric indices. In exactly similar fashion, we can take the 
remaining two terms on  the r.h.s. of (A.9)  and prove  that: 
\[
(\partial^0\,  \bar f^i - \partial^i\, \bar f^0)\, f_i - (\partial^0\, \bar  F^i - \partial^i\, \bar  F^0)\, F_i 
= [\partial^0\, (\bar f^i  + \bar F^i - \partial^i\,\bar C_1) 
\]
\[
- \partial^i\, (\bar f^0  + \bar F^0 - \partial^0\, \bar C_1)]\, f_i
- (\partial^0\, \bar  F^i - \partial^i\, \bar  F^0)\, ( f_i  +  F_i - \partial_i\,C_1). 
\eqno (A.18)
\]
Thus, ultimately, we have obtained the following   
\[
s_{ab}\, Q_b^{(1)} = \int d^{D-1}\, x \Big[ 
[\partial^0\, \{\bar f^i  + \bar F^i - \partial^i\, \bar C_1\}
- \partial^i\, (\bar f^0  + \bar F^0 - \partial^0\, \bar C_1)]\, f_i 
\]
\[
+ [\partial^0\, \{ f^i  +  F^i - \partial^i\, C_1\}
- \partial^i\, ( f^0  + F^0 - \partial^0\, C_1)]\, \bar f_i ~~~~
\]
\[
~~~~~- (\partial^0\, B^{ij} + \partial^i\, B^{j0} + \partial^j\, B^{0i})\, \{ B_{ij} 
+ \bar B_{ij} - (\partial_i\, \phi_j - \partial_j\, \phi_i)\}
\]
\[
~~~~~~~~~~~~~~~~~~~~- (\partial^0\, \bar F^i - \partial\, \bar F^0)\, (f_i + F_i - \partial_1\, C_1) 
- (\partial^0\, F^i - \partial^i\, F^0)\, (\bar f_i + \bar F_i - \partial_1\,\bar  C_1)
\]
\[
~~~~~~~~~~~~~~~~~~~~~~~~~~~~~+ \partial^0\, \{ B^{ij} + \bar B^{ij} - (\partial^i\, \phi^j - \partial^j\, \phi^i) \}B_{ij} 
+ \partial^i\; \{ B^{j0} + \bar B^{j0} - (\partial^j\, \phi^0
- \partial^0\, \phi^j) \}B_{ij} 
\]
\[
+ \;\partial^j\, \{ B^{0i} + \bar B^{0i} - (\partial^0\, \phi^i - \partial^i\, \phi^0) \}B_{ij} \Big], ~~~~~~~~~~~~~~~~~~
\eqno (A.19)
\]
which is equal to zero iff {\it all} the three CF-type restrictions: 
$B_{\mu\nu} + \bar B_{\mu\nu} - (\partial_\mu\, \phi_\nu - \partial_\nu\, \phi_\mu) = 0, \; 
f_\mu + F_\mu - \partial_\mu\, C_1 = 0, \; \bar f_\mu + \bar F_\mu - \partial_\mu\, \bar C_1 = 0$
are satisfied. In other words, we have been able to demonstrate the existence of the CF-type restrictions from the 
requirement of the absolute anticommutativity of the conserved and off-shell nilpotent versions of the
(anti-)BRST charges [cf. Eq. (A.1)].

We end this Appendix with the final remark that the absolute anticommtativity of the nilpotent (anti-)BRST charges 
$[Q_{(a)b}^{(1)}]$ can {\it also} be proven from
\[
s_{b}\, Q_{ab}^{(1)} = -\, i\, \{Q_{ab}^{(1)}, \, Q_{b}^{(1)} \} = 0 \Longrightarrow
Q_{ab}^{(1)}\, Q_{b}^{(1)} + Q_{b}^{(1)}\, Q_{ab}^{(1)} = 0, 
\eqno (A.20)
\]
where the l.h.s. can be computed by taking into account the BRST symmetry transformations (16) and 
the exploit expression for $Q_{ab}^{(1)}$ from (47). We would like to perform exactly similar kinds of 
computations as in the proof of $s_{ab}\, Q_{b}^{(1)} =  -\, i\, \{Q_{b}^{(1)}, \, Q_{ab}^{(1)} \} = 0 $
 where we have, ultimately, ended up with the expression (A.19). 
In fact we have accomplished this goal and the final answer is as follows
\[
s_b\, Q_{ab}^{(1)}   =   \int d^{D-1} x \, \Big[
(\partial^0\, \bar B^{ij} + \partial^i\, \bar B^{j0} + \partial^j\, \bar B^{0i})
\,\big\{ B_{ij} +  \bar B_{ij} - (\partial_i\, \phi_j - \partial_j\, \phi_i) \big\}
\]
\[
~~~~- \partial^0 \big\{ B^{ij} +  \bar B^{ij} - (\partial^i \phi^j - \partial^j \phi^i)   \big\}\,\bar B_{ij}
- \partial^i \big\{ B^{j0} +  \bar B^{jo} - (\partial^j \phi^0 - \partial^0 \phi^j)  \big\}\, \bar B_{ij} 
\]
\[
- \partial^j\, \big\{ B^{0i} +  \bar B^{0i} - (\partial^0\, \phi^i - \partial^i\, \phi^0)\,   \big\}\,\bar B_{ij}
- (\partial^0  F^i - \partial^i F^0) (\bar f_i + \bar F_i - \partial_i\bar C_1) 
\]
\[
- (\partial^0\,  \bar F^i - \partial^i\, \bar F^0)\, ( f_i +  F_i - \partial_i\,  C_1)
+ \partial^0\, \{\bar f^i + \bar F^i - \partial^ i\bar C_1 \}\, f_i ~~~~~~~~~~~~~~~~
\]
\[
~~~~~~~~~~~~~- \partial^i\, \{\bar f^0 + \bar F^0 - \partial^0 \bar C_1 \}\, f_i 
+  \partial^0\, \{ f^i +  F^i - \partial^ i C_1 \}\, \bar f_i 
- \partial^i\, \{ f^0 + F^0 - \partial^0  C_1 \}\, \bar f_i 
\Big], 
\eqno (A.21)
\]
which demonstrates that the r.h.s. can be zero if and only of {\it all} the {\it three} CF-type restrictions are satisfied.
Thus, the requirement of the absolute anticommutativity of the nilpotent (anti-)BRST charges  [$ Q_{(a)b}^{(1)}$] also leads to the 
existence of {\it three} CF-type restriction(s) on our theory which are {\it physical} in the sense that these are (anti-)BRST invariant.

\end{document}